\documentclass[12pt]{article}
\usepackage{epsfig}
\title{\bf Two-dimensional QCD in the Coulomb gauge}
\author{Yu.S.Kalashnikova\thanks{e-mail: yulia@heron.itep.ru}\hspace*{2mm}$^{\rm a}$, 
A.V.Nefediev\thanks{e-mail: nefediev@heron.itep.ru
nefediev@cfif.ist.utl.pt}\hspace*{2mm}$^{\rm a,b}$\\[5mm]
${\rm ^a}$ {\small\it Institute of Theoretical and Experimental Physics,}\\ 
{\small\it 117218, B.Cheremushkinskaya 25, Moscow, Russia}\\
${\rm ^b}$ {\small\it Grupo Te\'orico de Altas Energias (GTAE),}\\
{\small\it Centro de F\'\i sica das Interac\c c\~oes Fundamentais (CFIF),}\\
{\small\it Departamento de F\'\i sica, Instituto Superior T\'ecnico,} \\
{\small\it Av. Rovisco Pais, P-1049-001 Lisboa, Portugal}}
\date{}
\newcommand{\be}{\begin{equation}}
\newcommand{\ee}{\end{equation}}
\newcommand{\ra}{\rightarrow}
\newcommand{\too}{\mathop{\to}\limits_{N_C\to\infty}}
\newcommand{\tooo}{\mathop{\approx}\limits_{N_C\to\infty}}
\newcommand{\vpint}{\int\makebox[0mm][r]{\bf --\hspace*{0.13cm}}}
\newcommand{\ds}{\displaystyle}
\newcommand{\vph}{\varphi}
\newcommand{\trr}{\mathop{Tr}\limits_{x\to y+}}
\renewcommand{\theequation}{\thesection.\arabic{equation}}
\makeatletter
\@addtoreset{equation}{section}
\makeatother
\begin{document}
\maketitle

\begin{abstract}
In the present paper we discuss various aspects of the 't~Hooft model for
two-dimensional QCD in the limit of infinite number of colours in the Coulomb 
gauge. The properties of mesonic excitations are addressed, with special
attention paid to the pionic one. The twofold role of the pion is
discussed: being a genuine $q\bar q$ state it is also a Goldstone boson
of two-dimensional QCD. In particular, it is demonstrated explicitly
how the soft-pion theorems are satisfied. It is pointed out that the
Coulomb-gauge choice seems to be indispensable in studies of
hadronic observables with the pions involved.
\end{abstract}
\tableofcontents

\section{Introduction}
Two-dimensional QCD in the limit $N_C \rightarrow \infty$
(the 't~Hooft model \cite{'tHooft}) was first considered in 70-th. Since
then the
't~Hooft model is widely used as a toy laboratory for studies of
various aspects of strong interactions. 
The theory is relatively
simple, as there are no transverse gluons in two dimensions;
moreover, in the large-$N_C$ limit only planar graphs are to be summed up,
and the theory is exactly solvable.
Nevertheless, this is a truly relativistic field theory which does have a nontrivial
content,
resembling in such a way realistic QCD$_4$ case.
Indeed,
\begin{itemize}
\item the theory exhibits confinement and it is possible to 
demonstrate explicitly the existence of the discrete spectrum of the
quark-antiquark bound states;
\item the Poincar{\' e} invariance is maintained;
\item the chiral symmetry is spontaneously broken;
\item the Goldstone boson responsible for the chiral symmetry breaking is the $q\bar q$
ground state.
\end{itemize}

The first point is an almost trivial statement, since the Coulomb force is confining in
two dimensions.

The second item is of a paramount significance for hadronic spectroscopy. 
It was demonstrated explicitly in \cite{Bars&Green}
that, if a nonabelian theory is quantized in the explicitly noncovariant gauge,
a special care should be taken of the Lorentz invariance. The quantum Poincar{\' e}
algebra is closed only in the colour-singlet sector, which means that the spectrum can be
evaluated in an arbitrary frame including, for example, the centre-of-mass
frame as well as the infinite-momentum one. 

The chiral issue was historically a bit controversial.
The initial studies in QCD$_2$ were performed in the light-cone
gauge. The pioneering paper \cite{'tHooft} was followed by detailed
studies of spectra and wave functions of mesons as well as 
hadronic interactions \cite{CCG,Einhorn}. A bit later an
alternative approach was suggested, based on the Coulomb gauge $A_1=0$
\cite{Bars&Green}. The main advantage of the light-cone quantization
is considerable simplification of the spectra calculations, but 
straightforward analysis gives the perturbative vacuum. The more technically
involved version \cite{Bars&Green} yields a nontrivial vacuum, and it
appears that a
nonzero quark condensate exists for the massless quarks \cite{Ming Li}.
The latter feature is confirmed by the sum rule calculations in the
light-cone gauge \cite{Zhitnitskii,Burkardt}. At the hadronic level 
the apparent discrepancies are connected with a peculiar
form of the pionic wave function near the chiral limit in the light-cone
quantization, as
it is discussed in \cite{Zhitnitskii}, and will be explicitly
demonstrated below. That is why one is forced to employ the sum rules and
the Operator Product Expansion (OPE) to arrive at reliable results in the 
pionic physics and the vacuum structure
\cite{Zhitnitskii}. On the contrary, the choice of the Coulomb gauge
does not lead to drastic singularities and enables treating the pions on the
same footing as other mesons. The conceptual difficulties of the
light-cone quantization were
resolved in the formulation on finite intervals \cite{Lenz}, where
the equivalence of both versions was demonstrated explicitly, clarifying
the relationship between the light-cone and the equal-time quantization. 

Since then a lot of work in two-dimensional QCD was done, employing the
light-cone gauge. Among the questions discussed are the general properties
of the OPE \cite{Zhitnitskii, Shifman} and heavy quark expansion and
duality \cite{Uraltsev}. The calculations of spectra were performed in the
framework of the so-called discretized light-cone quantization beyond the
$N_C\to\infty$ limit \cite{Hornbostel}. A separate fascinating
issue is the studies of QCD$_2$ with adjoint fermions \cite{Klebanov}.
In the present paper we discuss the properties of vacuum and low-lying
mesonic states built of light quarks in the Coulomb gauge, with special
attention paid to the chiral issues of the theory.
  
Quantization on the light-cone allows one to establish an obvious connection
with the dynamics of the parton model, while quantization on the ordinary
time hypersurface is natural for another important branch of
phenomenology, the constituent quark model. Indeed, the spectrum of QCD$_2$ is
discrete, and $N_C \rightarrow \infty$ limit suppresses additional quark
pair creation, so that the 't~Hooft model is nothing but a constituent
quark model exactly derived from a nontrivial relativistic quantum 
field theory.

In the constituent quark models the confinement is usually modeled by
a potential force. Then the gross features of the light quarkonia spectra
and decays are described surprisingly well with an exception of the pion.
Since one cannot include chiral symmetry breaking (CSB) effects into the
constituent picture, there is no hope to reproduce the pion as the Goldstone
boson, and soft pion theorems cannot be satisfied within any naive quark
model picture.

The CSB phenomena follow from the most general symmetry considerations and
have nothing to do with the particular mechanism of the confinement.
One possible way to include the soft pions into the quark model is to
introduce quarks and pions on equal footing, as it is done in the
chiral quark models (see, {\it e.g.}, \cite{Vento} and references therein). 
Nevertheless, there are
some results from the lattice simulations \cite{lattice} demonstrating 
that the confinement and the CSB phenomena are present in the confining 
phase and disappear in unison above the
deconfinement temperature; so both phenomena are interrelated dynamically.

A model was suggested many years ago \cite{Orsay}, which connects 
the confinement and CSB (see also \cite{port}, where similar ideas were 
employed). The main ingredient of this model is an instantaneous
three-dimensional oscillator confining force. Such a force does not follow
from QCD and, in addition, there is no gauge and Lorentz invariance
in this approach. An important development was suggested in
\cite{Sim2}, where the QCD vacuum is parametrized by a set of
gauge and Lorentz invariant nonperturbative gluonic correlators which
are responsible for both, the area law and the chiral condensate formation.
The quark model which follows from such an approach should be able to
reproduce, {\it inter alia}, all pion properties. In this regard it is 
instructive to study an exactly solvable theory with confinement and
CSB, and the 't~Hooft model is a perfect toy laboratory for such studies.

Before proceeding further we would like to note that the large-$N_C$ limit
is essential in establishing the chiral properties of QCD$_2$ \cite{Witten}.
The Coleman theorem \cite{Coleman} prohibits CSB for any finite number
of degrees of freedom in a two-dimensional theory. It means that limits $N_C
\rightarrow \infty$ and $m_q \rightarrow 0$ are not interchangeable.
There is no contradiction with the Coleman theorem if one considers the
weak coupling regime where $m_q\gg g\sim 1/\sqrt{N_C}$, {\it i.e.}, the limit
of infinite number of colours is taken first (see \cite{Zhitnitskii}
for the detailed discussion of this issue as well as of the other phase of
the theory which corresponds to the strong coupling regime $m_q\ll g$).
 
The theory is given by the Lagrangian
\be
L (x)= -\frac{1}{4}F^a_{\mu\nu}(x)F^a_{\mu\nu}(x) + \bar q(x)(i\hat{D} - m)q(x),
\label{lagrangian}
\ee
$$
\hat{D} = (\partial_{\mu} - igA^a_{\mu}t^a)\gamma_{\mu},
$$
and the large-$N_C$ limit implies that
\be
\gamma=\frac{g^2N_C}{4\pi}\too {\rm const}.
\ee

Our convention for $\gamma$ matrices is
\be
\gamma_0\equiv\beta=\sigma_3,~\gamma_1=i\sigma_2,~\gamma_5\equiv\alpha=\gamma_0\gamma_1=\sigma_1.
\ee
The theory is gauged by the condition
\be
A_1(x_0,x)=0,
\ee
so that the only nontrivial correlator of the gluonic fields, the gluon propagator, 
takes the form
$$
D^{ab}_{01}(x_0 - y_0, x - y )=D^{ab}_{11}=(x_0 - y_0, x - y )=0,
$$
\be
D^{ab}_{00}(x_0 - y_0, x - y ) =-\frac{i}{2}\delta^{ab} |x-y|\delta(x_0 -
y_0),
\label{D}
\ee
and the infrared singularity is regularized by the principal value
prescription, {\it i.e.}, by means of an appropriate number of subtractions, for
example,
$$
\int\frac{dk}{(p-k)^2}F(k)\to\vpint\frac{dk}{(p-k)^2}F(k)=\hspace*{5cm}
$$
\be
\hspace*{3cm}=\int\frac{dk}{(p-k)^2}\left(F(k)-F(p)-F'(p)(k-p)-\ldots\right),
\label{regpres}
\ee
yielding a linear confinement for the interquark interaction mediated by the two-dimensional
gluon.

The paper is organized as follows. In Section 2 a Hamiltonian approach
to QCD$_2$ in the Coulomb gauge is developed. The bosonization of the model is
performed explicitly in the large-$N_C$ limit, and a generalized
Bogoliubov-Valatin transformation is employed to construct the 
composite operators creating and annihilating bosons. The pion wave function
is found explicitly and the chiral properties of the theory are discussed.
In Section 3 a matrix formalism is presented, a matrix bound-state 
equation is derived with special attention paid to the problem of the
truncated Hilbert space and Hermiticity. We clarify the role of the backward
motion of the $q \bar q$ pair in the meson. 
The current conservation and the Ward 
identities are discussed as well as the pionic vertex. Section 4 is devoted to
strong hadronic decays, where we check the low-energy theorems, including the
\lq\lq Adler zero" selfconsistency condition. 
Concluding remarks are given in Section 5.

\begin{figure}[t]
\begin{center}
\epsfxsize=14cm
\epsfbox{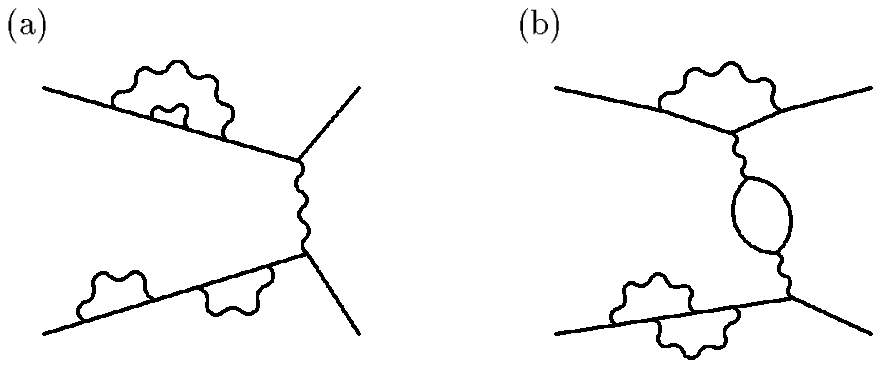}
\caption{The planar (figure (a)) and the nonplanar, suppressed by $N_C$, (figure (b))
diagrams.}\label{figdiag}
\end{center}
\end{figure}

\section{Hamiltonian approach}

As is known from the pioneering work \cite{'tHooft}, the physical degrees of freedom
in two-dimensional QCD in the weak-coupling regime are $q\bar q$ mesons. It would
be quite natural then to reformulate the model entirely in terms of compound
mesonic states, introducing a nonperturbative vacuum and creation and annihilation 
operators for the mesons, as excitations over this vacuum. The most natural framework
for such a task is the Hamiltonian approach to the model, which we develop in this
section.\footnote{The Hamiltonian approach to two-dimensional QCD
in the light-cone gauge was developed in \cite{japan}.} Note that it
is possible in a selfconsistent
form due to an instantaneous type of the interaction induced by (\ref{D}).

This section is organized as follows. In subsection \ref{dressq} we start from the
Hamiltonian of the model in the Coulomb gauge, introduce dressed quark fields and,
following \cite{Bars&Green}, derive a gap equation (also known as mass-gap equation). 
We discuss the numerical 
solution to the gap equation found in \cite{Ming Li}. Subsection \ref{twovac} is devoted
to investigation of the vacuum energy and identification of the chirally-symmetric and
nonsymmetric vacua. In subsection \ref{genbog}
we introduce two-particle operators and perform a generalized Bololiubov
transformation to diagonalize the Hamiltonian of the model in the mesonic sector.
In subsection \ref{bseq} we discuss properties and solutions of the bound-state equation,
which
appears as a consequence of the second Bogoliubov transformation.
A special solution to the bound-state equation, the massless chiral pion, is found
analytically
and investigated in subsection \ref{pn}. An issue connected to the locality and the 
Lorentz nature of
confining interaction in two-dimensional QCD is discussed in subsections \ref{lnc} and
\ref{qqq}. In
conclusion we turn to the chiral properties of the model and this is the subject of 
subsection \ref{cpm}.

\subsection{Dressed quarks and the mass-gap equation}
\label{dressq}

Starting from the Lagrangian (\ref{lagrangian}) and following the standard rules one
arrives at the Hamiltonian of the theory in the form
$$
H=\int dxq^{+}(t,x)\left(-i\gamma_5\frac{\partial}{\partial x}+m\gamma_0\right) 
q(t,x)-\hspace*{3cm}
$$
\be
\hspace*{3cm}
-\frac{g^2}{2}\int\int
dxdy\;q^{+}(t,x)t^aq(t,x)q^{+}(t,y)t^aq(t,y)\frac{\left|x-y\right|}{2}.
\label{H}
\ee

Note that only four-quark interaction enters the Hamiltonian (\ref{H}) and this is
a reflection of the fact stated in the Introduction that the only nontrivial
gluonic correlator is the gluonic propagator (\ref{D}). In
four-dimensional QCD, possessing a much more complicated vacuum structure,
correlators of all orders should appear, which gives rise to
an infinite number of terms with four-quark, six-quark, 
{\it etc} interactions in the Hamiltonian. In the meantime, 
it is reasonable to truncate the QCD$_4$ 
Hamiltonian at the four-quark interaction level, which corresponds to the
bilocal approximation, when only $\langle AA\rangle$ correlator is 
left.\footnote{These are irreducible averages (cumulants) meant here, which
are defined as $\langle\langle O\rangle\rangle=\langle O\rangle$,
$\langle\langle O_1O_2\rangle\rangle=\langle O_1O_2\rangle-\langle
O_1\rangle\langle O_2\rangle$, {\it etc} \cite{cumul}. Due to the Lorentz and the colour
invariances of the QCD vacuum $\langle A_{\mu}^a\rangle=0$ and, hence,
$\langle\langle A_{\mu}^aA_{\nu}^b\rangle\rangle=\langle
A_{\mu}^aA_{\nu}^b\rangle$.} The interested reader can find a detailed review 
of the given approach in \cite{YuArev}. 

The \lq\lq dressed" quark field $q(x)$ in (\ref{H}) is defined as follows
\cite{Bars&Green}
\be
q_{\alpha i}(t,x)=\int\frac{dk}{2\pi}e^{ikx}[b_{\alpha}(k,t)u_i(k)+d_{\alpha}^+
(-k,t)v_i(-k)],
\label{quark_field}
\ee
\be
b_{\alpha}(t,k)|0\rangle= d_{\alpha}(t,-k)|0\rangle =0,\quad
b^{+}_{\alpha}(t,k)|0\rangle=|q\rangle,\quad
d^{+}_{\alpha}(t,-k)|0\rangle=|\bar{q}\rangle,
\label{bnd}
\ee
\be
\ds\{b_{\alpha}(t,p)b^+_{\beta}(t,q)\}=
\ds\{d_{\alpha}(t,-p)d^+_{\beta}(t,-q)\}=2\pi\delta(p-q)\delta_{\alpha\beta},
\label{bdcommutators}
\ee
\be
u(k)=T(k)\left(1 \atop 0 \right),\quad v(-k)=T(k)\left(0 \atop 1 \right),
\label{unv}
\ee
$$
T(k)=e^{-\frac{1}{2}\theta(k)\gamma_1},
$$
where $\theta$ plays the role of the Bogoliubov-Valatin angle.
Greek and Latin letters denote colour and spinor indices, respectively.
Strong interaction between quarks implies that the true vacuum state 
is described by a nontrivial $\theta$, whereas excitations over it bring positive
contribution to the energy.

In what follows we shall omit the explicit dependence of operators on time.
It can be easily restored at any intermediate step, thus giving, for example,
\be
b_{\alpha}(t,p)=b_{\alpha}(p)e^{-iE(p)t},\quad
d_{\alpha}(t,-p)=d_{\alpha}(-p)e^{iE(p)t},
\ee
where $E(p)$ is the dispersive law of the dressed quark (to be defined later).

The Hamiltonian (\ref{H}) normally ordered in the basis (\ref{bnd}) splits into
three parts ($L$ is the one-dimensional volume),
\be
H= LN_C{\cal E}_v + :H_2: + :H_4:,
\label{Hh}
\ee
where
\be
{\cal E}_v =\int\frac{dp}{2\pi}Tr
\left\{\left(\gamma_5p+m\gamma_0\right)\Lambda_{-}(p)+\frac{\gamma}{4\pi}
\int\frac{dk}{(p-k)^2}\Lambda_{+}(k)\Lambda_{-}(p)\right\}
\label{vac}
\ee
is the vacuum energy density,
$$
:H_2:=\int dx:q^{+}_i(x)\left(-i\gamma_5\frac{\partial}{\partial
x}+m\gamma_0\right)q_i(x):-
\hspace*{5cm}
$$
\be
-\frac{\gamma}{2}\int dxdy\frac{|x-y|}{2}\int dk :q^{+}_i(x)
\left[\Lambda_{+}(k)-\Lambda_{-}(k)\right]q_i(y):e^{ik(x-y)}
\label{H2}
\ee
is quadratic in quark fields, whereas the $:H_4:$ part contains four of them,
\be
:H_4:=-\frac{g^2}{2}\int dxdy
:q^{+}(x)t^aq(x)q^{+}(y)t^aq(y):\frac{\left|x-y\right|}{2}.
\label{H4}
\ee
We have introduced projectors onto positive and negative states for convenience:
\be
\Lambda_{\pm}(p)=T(p)\frac{1\pm\gamma_0}{2}T^+(p).
\label{projectors}
\ee

One comment on the role played by different parts of the Hamiltonian (\ref{Hh}) is
in order. The first term in (\ref{Hh}) defines the energy of the vacuum which
is to be minimized. The
$:H_2:$ part describes \lq\lq dressing" of quarks, so that an alternative approach to
minimizing ${\cal E}_v$ is the requirement that $:H_2:$ be diagonal in terms of
the quark creation and annihilation operators, or, equivalently, that the 
anomalous
Bogoliubov term be absent. No matter which way is used, the resulting equations
for the Bogoliubov-Valatin angle $\theta$ and the dispersive law for the dressed
quarks read \cite{Bars&Green}:
\be
\left\{
\begin{array}{c}
E(p)\cos\theta(p)=m+\frac{\ds \gamma}{\ds 2}\ds\vpint\frac{dk}{(p-k)^2}
\cos\theta(k)\;\\
{}\\
E(p)\sin\theta(p)=p+\frac{\ds \gamma}{\ds 2}\ds\vpint\frac{dk}{(p-k)^2}
\sin\theta(k),
\end{array}
\right.
\label{system}
\ee
which can be reformulated in the form of the gap equation for the angle $\theta$,
\be
p\cos\theta(p)-m\sin\theta(p)=\frac{\gamma}{2}\vpint\frac{dk}{(p-k)^2}\sin[\theta(p)-\theta(k)].
\label{gap}
\ee

If a solution for $\theta$ is known, $E(p)$ can be easily found from the relation
\be
E(p)=m\cos\theta(p)+p\sin\theta(p)+\frac{\gamma}{2}\vpint\frac{dk}{(p-k)^2}
\cos[\theta(p)-\theta(k)].
\label{E}
\ee

\begin{figure}[t]
\centerline{\epsfig{file=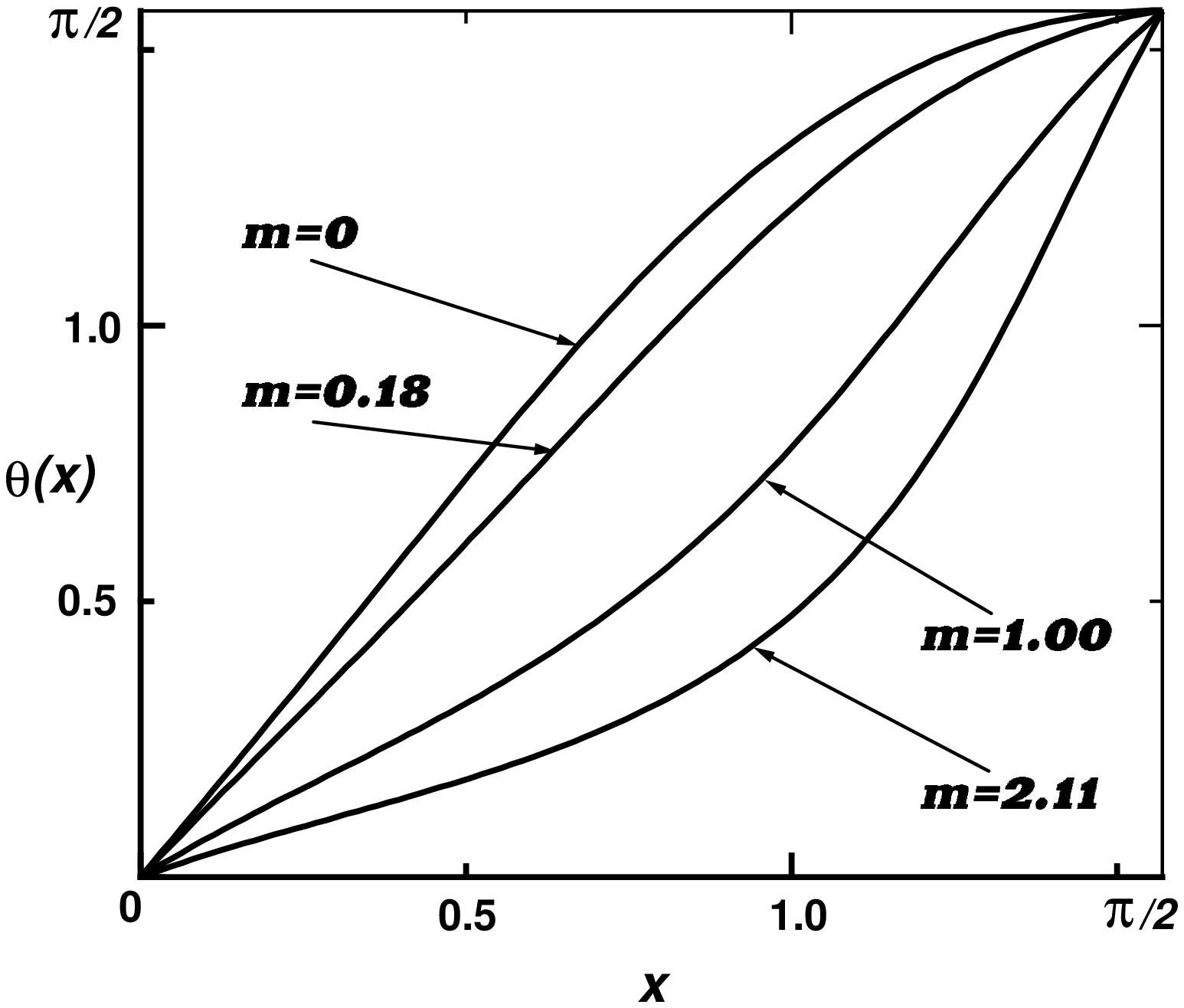,width=10cm}\hspace{-1.8cm}
            \epsfig{file=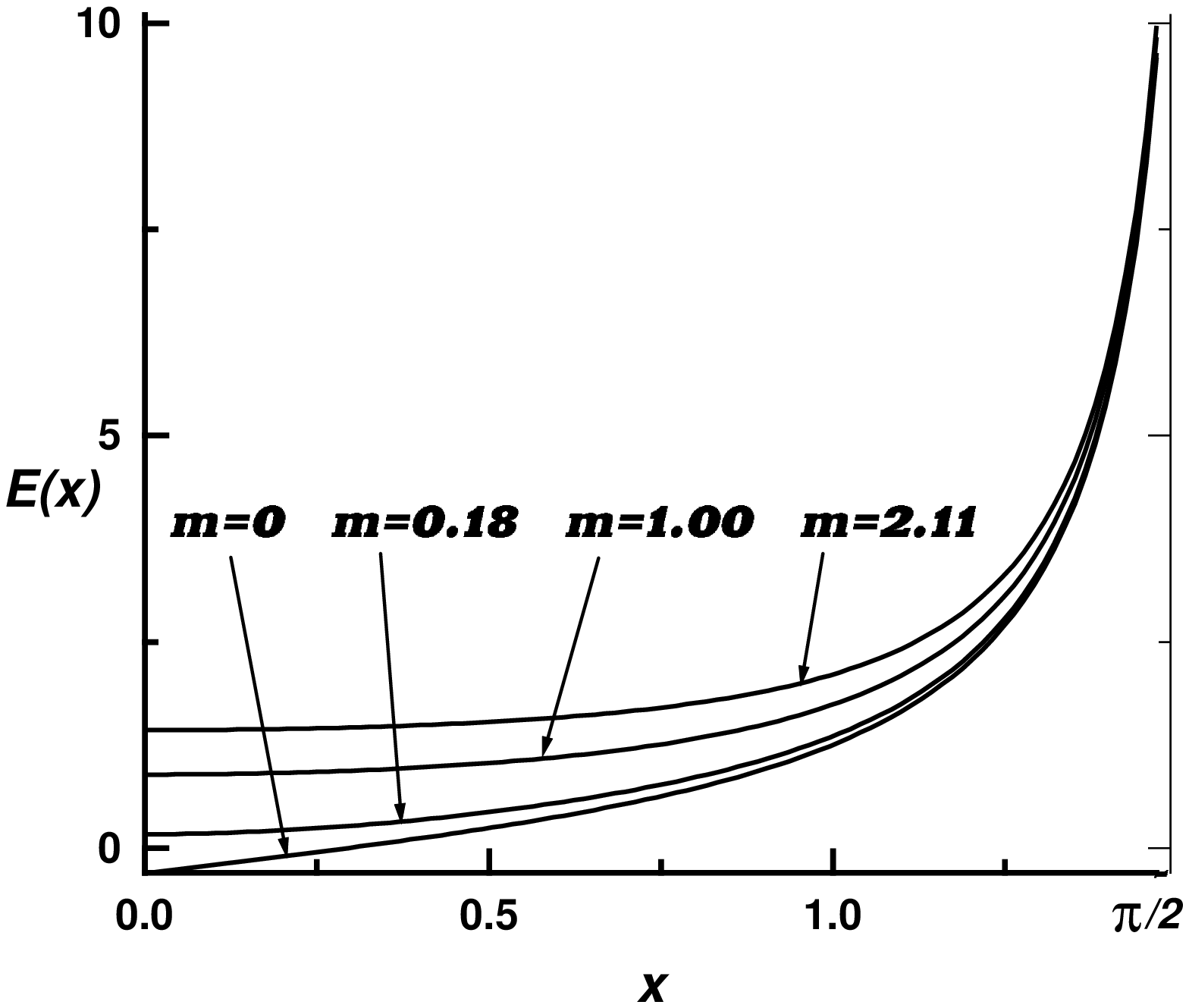,width=10cm}}
\caption{Numerical solutions for the Bogoliubov angle $\theta(p)$ and the
dressed quark dispersive law $E(p)$ for different values of the quark
mass. The plots are taken from \cite{Ming Li}, where $x$ comes from the
change of variable $p=\tan(x)$, all dimensional quantities are given in the proper 
units of $(2\gamma)^{1/2}$.}\label{figtheta}
\end{figure}

The gap equation (\ref{gap}) is subject to numerical investigation which was
performed in \cite{Ming Li}. The results for $\theta(p)$ and $E(p)$ are given in
Fig.\ref{figtheta}.  

Notice several important properties of the functions $\theta$ and $E$:
\begin{itemize}
\item $\theta(p)$ is odd, whereas $E(p)$ is even,
\be
\theta(-p)=-\theta(p),\quad E(-p)=E(p);
\ee
\item solution for $\theta(p)$ remains nontrivial even in the chiral 
limit, $m=0$, and so does the solution for 
the dispersive law (see Fig.\ref{figtheta});
\item the function $E(p)$ is not positively defined, as it might be naively expected 
(see Fig.\ref{figtheta}).
\end{itemize}

The formal reason for the last property comes from the regularization prescription 
(\ref{regpres}), which defines the way of subtracting an infinite self-energy constant.
Thus a divergent integral of positively defined functions
might lead to a negative result after the proper regularization. 

One can easily check that 
\be
\theta(p\to 0)\sim p,\quad \theta'(0)>0,
\ee
and
\be
E(0)=m+\frac{\gamma}{2}\vpint\frac{dk}{k^2}\cos\theta(k)=
m-\frac{\gamma}{2}\int_0^{\infty}\frac{dk}{k^2}\sin^2\frac{\theta(k)}{2}\approx 
m-\frac{\pi\gamma}{8}\theta'(0),
\ee
so that $E(0)$ becomes negative for $m$ smaller than some critical value $m_c$.
This does not lead to a disaster as, according to findings of \cite{Bars&Green}, only
the colour-singlet sector of the theory is Lorentz covariant, whereas the quark
sector is not colour-singlet and, hence, it may not be Lorentz covariant.
Nevertheless, one has to be especially careful shifting the poles
in the quark propagator, paying attention to the sign of the product of $E(p)$
and the infinitely small positive constant $\varepsilon$
\cite{Bars&Green}\label{pageE}. 

Note that a simple analytical solution to the gap
equation (\ref{gap}) in the chiral limit was found in \cite{Bars&Green}, which 
reads\footnote{In fact, any odd function $\theta(p)$ with an arbitrary number of
jumps from $\pi/2$ to $-\pi/2$ and back also satisfies the gap equation (\ref{gap})
\cite{Bars&Green}. Meanwhile solutions of such a type do not reduce to the free
theory when $\gamma$ tends to zero and they will not be discussed.}
\be
\theta(p)=\frac12\pi {\rm sign}(p),\quad E(p)=|p|-P\frac{\gamma}{|p|},
\label{badsol}
\ee
where the symbol $P$ stands for the principal value prescription.
It is clearly seen from (\ref{badsol}) that for this solution 
$E(p)$ is not positively defined either. 

The two solutions to the gap equation in the chiral limit, the one given by
(\ref{badsol}) and the other, depicted in Fig.\ref{figtheta}, define two
different phases of the theory. The chiral symmetry is broken in the latter case
only, whereas in the former one it remains unbroken. 

It was demonstrated in \cite{BNR} that the analytical solution (\ref{badsol})
possesses infinite energy compared to the numerical chirally-nonsymmetric one given
in Fig.\ref{figtheta}. Thus the chiral symmetry is never restored in the
't~Hooft model and there is only one phase of the theory with chiral symmetry
spontaneously broken (see the next subsection for the details). Such way, the solution (\ref{badsol}) is unphysical, but we
still prefer to keep it to exemplify some statements concerning the chiral
properties of the model. In the next subsection we discuss this issue in more detail.

Performing all necessary calculations with the precautions discussed above, one
arrives at the Hamiltonian of the model which is diagonal with respect to
the dressed quarks basis. The contribution of the $:H_4:$ part of the
Hamiltonian (\ref{Hh}) is suppressed by an extra factor $1/\sqrt{N_C}$ and thus it
can be neglected in the single-quark sector of the theory;
\be
:H:=LN_C{\cal E}_v+\int\frac{dk}{2\pi}E(k)\left\{b_{\alpha}^{+}(k)b_{\alpha}(k)+
d_{\alpha}^{+}(-k)d_{\alpha}(-k)\right\}.
\label{Hq}
\ee

Note that the result (\ref{Hq}) itself has practically no value as it deals with
the gauge- and Lorentz-noncovariant sector.
The most important result of this subsection is the gap equation
(\ref{gap}) and the numerical solution to it depicted in Fig.\ref{figtheta}. 
They will be intensively used in what follows.

\subsection{The vacuum energy. Chirally-symmetric and nonsymmetric va\-cua}\label{twovac}

Let us return to the gap equation (\ref{gap}) and discuss an
alternative way of its derivation --- namely, the minimization of the vacuum energy
(\ref{vac}) \cite{BNR}. We consider the case of massless quarks, $m=0$.

It is convenient to introduce an excess of the vacuum energy density for the theory with
interaction over the free-theory one,
\be
\Delta{\cal E}_v[\theta]={\cal E}_v[\theta]-{\cal E}_v[\theta_{\rm free}]=-\int\frac{dp}{2\pi}(p\sin\theta(p)-|p|)
-\frac{\gamma}{4\pi}\int\frac{dpdk}{(p-k)^2}\cos[\theta(p)-\theta(k)],
\label{evac}
\ee
where $\theta_{\rm free}(p)=\frac{\pi}{2}{\rm sign}(p)$ corresponds to the free ($\gamma=0$)
massless theory.

The gap equation, following from the minimization procedure,
\be
\frac{\delta}{\delta\theta(p)}\Delta{\cal E}_v[\theta]=0,
\label{massgap}
\ee 
readily reproduces equation (\ref{gap}). 

To ensure that the solution to equation (\ref{gap}) indeed minimizes the vacuum energy
we use the following approach \cite{BNR}. 
Let $\theta(p)$ be the solution to (\ref{massgap}) corresponding to the minimum of
$\Delta {\cal E}_v$. Then $\theta(p/A)$ stretched with an arbitrary parameter
$A$ should enlarge the energy (\ref{evac}). Naive dimensional analysis demonstrates
that $\Delta {\cal E}_v$ scales with $A$ as
\be
\Delta{\cal E}_v=\frac12C_1A^2-\gamma C_2,
\label{EA}
\ee
with $C_{1,2}$ being positive constants. Then the stable solution is provided by
$A_0$ minimizing the energy (\ref{EA}), {\it i.e.}, by $A_0=0$, which corresponds 
either to the free massless theory, giving $\Delta {\cal E}_v=0$, or to the analytic
solution (\ref{badsol}). Both solutions correspond to the preserved chiral symmetry.
Thus one arrives at a discouraging conclusion that there is
no nontrivial chirally-nonsymmetric solution to the gap equation (\ref{gap}). 
In the meantime, the naive analysis performed above completely ignores the fact
that the vacuum energy (\ref{evac}) is logarithmically infrared divergent due to the 
second term. Introducing a cut-off and repeating the same steps, one can conclude that the
correct dependence of the vacuum energy on the scale parameter $A$ is as follows:
\be
\Delta{\cal E}_v=\frac12C_1A^2-\gamma C_2\ln A+\gamma C_3
\label{EA2}
\ee
instead of the naive form (\ref{EA}). The constant $C_3$ is proportional to the logarithm
of the cut-off and can be removed by an infinite renormalization. 

Note that the second term in (\ref{evac}) cannot be made convergent both, in the infrared
and in the ultraviolet simultaneously. Indeed, one can remove the infrared divergence in 
(\ref{evac}),
subtracting the energy corresponding to the solution (\ref{badsol}) instead of the free
one. Then the resulting integral appears ultraviolet logarithmically divergent and leads
to the same relation (\ref{EA2}) but with the infinite constant $C_3$ containing the
logarithm of the ultraviolet cut-off.

The function (\ref{EA2}) always has a minimum at
\be
A_0=\sqrt{\gamma\frac{C_2}{C_1}},
\label{A0}
\ee
which corresponds to the nontrivial solution of the gap equation found
numerically in \cite{Ming Li} and depicted in Fig.\ref{figtheta}. 
In the meantime, from the form of the function (\ref{EA2}) one
can see the logarithmic growth of the energy in approaching the solution (\ref{badsol}),
which, as was discussed above, corresponds to $A_0=0$. 

Thus ones arrives at the conclusion, already mentioned above, 
that the vacuum energy corresponding to the
chirally-symmetric solution (\ref{badsol}) is infinite compared to that for the
chirally-nonsymmetric one depicted in Fig.\ref{figtheta}\footnote{Once 
the logarithmically divergent term in 
(\ref{EA2}) is proportional to the coupling constant $\gamma$, then there
is no problem with the free limit of the theory, which also corresponds to $A_0=0$.
Indeed, when $\gamma$ tends to zero the logarithmic term in (\ref{EA2}) disappears, so
that, as defined by (\ref{evac}), the vacuum energy of the free theory is zero.} and,
hence, no phase transition of the chiral symmetry restoration is possible in the 't~Hooft model.

It is instructive to note that the very fact of the existence of the
chirally-nonsymmetric nontrivial solution to the gap equation (\ref{gap}) is yet another 
consequence of the infrared behaviour of the 't~Hooft model discussed above. Indeed, the
gap equation (\ref{gap}) was derived in neglection of all effects of the fermionic vacuum
polarization (creation and annihilation fermionic operators introduced in (\ref{bnd})
correspond to the so-called BCS approximation). In the meantime, the chiral symmetry can be
spontaneously broken in other two-dimensional models for QCD, like the Gross-Neveu one
\cite{GN,Remb}, but in order to have a nonzero chiral condensate one has to go beyond 
BCS level, summing up fermionic bubbles, whereas
the equation similar to (\ref{gap}) has only a trivial chirally-symmetric solution, giving
$\langle \bar{q}q\rangle=0$. 

\subsection{Generalized Bogoliubov transformation and mesonic com\-po\-und states}
\label{genbog}

In the previous subsection the first two terms of the Hamiltonian (\ref{Hh})
were considered. 
Let us turn to the third part, $:H_4:$. With substituted 
solution for the dressed quarks, it describes interaction between them, which
leads to formation of the $q\bar q$ bound states --- mesons. 
In this subsection we are to diagonalize the Hamiltonian
(\ref{Hh}) in the colour-singlet mesonic sector of the theory. To this end we
introduce compound operators which act on colourless pairs of quarks and antiquarks
\cite{Lenz,KNV}:
\be
\begin{array}{c}
B(p,p')=\frac{\ds 1}{\ds\sqrt{N_C}}b_{\alpha}^{+}(p)b_{\alpha}(p'),\quad 
D(p,p')=\frac{\ds 1}{\ds\sqrt{N_C}}d_{\alpha}^{+}(-p)d_{\alpha}(-p'),\\
{}\\
M(p,p')=\frac{\ds 1}{\ds\sqrt{N_C}}d_{\alpha}(-p)b_{\alpha}(p'),\quad
M^{+}(p,p')=\frac{\ds 1}{\ds\sqrt{N_C}}b^{+}_{\alpha}(p')d^{+}_{\alpha}(-p),
\end{array}
\label{operators}
\ee
with the commutation relations being
$$
[M(p,p')M^{+}(q,q')]=(2\pi)^2\delta(p'-q')\delta(p-q)-
$$
\be
-\frac{2\pi}{\sqrt{N_C}}\left\{
D(q,p)\delta(p'-q')+B(q',p')\delta(p-q)\right\}\too
\label{mcom}
\ee
$$
\to(2\pi)^2\delta(p'-q')\delta(p-q),
$$
$$
\begin{array}{c}
[B(p,p')B(q,q')]=\frac{\ds 2\pi}{\ds\sqrt{N_C}}\left(B(p,q^{\prime})\delta(p'-q)-
B(q,p')\delta(p-q')\right)\too 0,\\
{}\\

[D(p,p')D(q,q')]=\frac{\ds 2\pi}{\ds\sqrt{N_C}}\left(D(p,q^{\prime})\delta(p'-q)-
D(q,p')\delta(p-q')\right)\too 0.\\
{}\\

[B(p,p')M(q,q')]=-\frac{\ds 2\pi}{\ds\sqrt{N_C}}M(q,p')\delta(p-q')\too 0,\\
{}\\

[B(p,p')M^+(q,q')]=\frac{\ds 2\pi}{\ds\sqrt{N_C}}M^+(q,p)\delta(p'-q')\too 0,\\
{}\\

[D(p,p')M(q,q')]=-\frac{\ds 2\pi}{\ds\sqrt{N_C}}M(p',q')\delta(p-q)\too 0,\\
{}\\

[D(p,p')M^+(q,q')]=\frac{\ds 2\pi}{\ds\sqrt{N_C}}M^+(p,q')\delta(p'-q)\too 0.\\
\end{array}
$$

With the new operators substituted, the Hamiltonian (\ref{Hh}) takes the form
$$
H=LN_C{\cal E}_v+\sqrt{N_C}\int\frac{dk}{2\pi}E(k)\{B(k,k)+D(k,k)\}
$$
$$
-\frac{\gamma}{2}\int\frac{dp\;dk\;dQ}{(2\pi)^2(p-k)^2}
\left[2\cos\frac{\theta(p)-\theta(k)}{2}\sin\frac{\theta(Q-p)-\theta(Q-k)}{2} 
\right.
$$
$$
\times\left\{M^{+}(p,p-Q)D(k-Q,k)+M^{+}(p-Q,p)B(k-Q,k)\right.\hspace*{1cm}
$$
$$
\hspace*{1cm}\left. -B(p,p-Q)M(k-Q,k)-D(p,p-Q)M(k,k-Q)\vphantom{M^+}\right\}
$$
\be
+\cos\frac{\theta(p)-\theta(k)}{2}\cos\frac{\theta(Q-p)-\theta(Q-k)}{2}
\label{HH}
\ee
$$
\times\left\{\vphantom{M^+}B(p-Q,p)B(k,k-Q)+D(p,p-Q)D(k-Q,k)\right.\hspace*{1cm}
$$
$$
\hspace*{1cm}+\left.M^{+}(p-Q,p)M(k-Q,k)+M^{+}(p,p-Q)M(k,k-Q)\right\}
$$
$$
+\sin\frac{\theta(p)-\theta(k)}{2}\sin\frac{\theta(Q-p)-\theta(Q-k)}{2}
$$
$$
\times\left\{\vphantom{M^+}B(p,p-Q)D(k,k-Q)+B(p-Q,p)D(k-Q,k)\right.\hspace*{1cm}
$$
$$
\left.\hspace*{1cm}\left.+M(p,p-Q)M(k-Q,k)+M^{+}(p-Q,p)M^{+}(k,k-Q)\right\}
\vphantom{\frac{dk}{(p-k)^2}}\right],
$$
where $:H_2:$ and $:H_4:$ terms should have the same order in powers of $N_C$ and thus
act together as opposed to the one-body sector, where $:H_4:$ was suppressed as
$1/\sqrt{N_C}$.

A crucial step we are to perform now is to note, that in the mesonic sector of
the theory one cannot create and annihilate isolated quarks and antiquarks.
Only colourless $q\bar q$ pairs can appear, so that, creating a quark, we have to create
an accompanying antiquark and, {\it vice versa}, for each created antiquark we have an extra
quark. Thus the operators
(\ref{operators}) cannot be independent. Indeed, it is easy to check that the
substitution
\be
\begin{array}{c}
B(p,p')=\frac{\ds 1}{\ds\sqrt{N_C}}\ds\int\frac{\ds dq''}{\ds 2\pi}M^{+}(q'',p)M(q'',p'),\\
{}\\
D(p,p')=\frac{\ds 1}{\ds\sqrt{N_C}}\ds\int\frac{\ds dq''}{\ds
2\pi}M^{+}(p,q'')M(p',q'')
\end{array}
\label{anzatz}
\ee
satisfies the commutation relations (\ref{mcom}). Now one can neglect a number 
of terms in the Hamiltonian (\ref{HH}) and to arrive at a simplified expression,
$$
H=LN_C{\cal E}_v+\int\frac{dQdp}{(2\pi)^2}\left[(E(p)+E(Q-p))M^{+}(p-Q,p)M(p-Q,p)\right.
$$
\be
-\frac{\gamma}{2}\int\frac{dk}{(p-k)^2}\left\{2C(p,k,Q)M^{+}(p-Q,p)M(k-Q,k)\right.
\label{HHH}
\ee
$$
\left.\left.+S(p,k,Q)\left(M(p,p-Q)M(k-Q,k)+M^{+}(p,p-Q)M^{+}(k-Q,k)\right)\right\}\right],
$$
where
\be
\begin{array}{c}
C(p,k,Q)=\ds\cos\frac{\ds\theta(p)-\theta(k)}{\ds 2}\cos\frac{\ds
\theta(Q-p)-\theta(Q-k)}{\ds2}\nonumber,\\
\vphantom{.}\\
S(p,k,Q)=\ds\sin\frac{\ds\theta(p)-\theta(k)}{\ds
2}\sin\frac{\ds\theta(Q-p)-\theta(Q-k)}{\ds2}\nonumber.
\label{cs}
\end{array}
\ee

The operators $M^+$ and
$M$ create and annihilate quark-antiquark pairs, which are not mesons yet since
Hamiltonian (\ref{HHH}) is not diagonal in terms of these operators. 

Symbolically the operator structure of the Hamiltonian (\ref{HHH}), 
\be
H\sim H_0+AM^+M+\frac12B(M^+M^++MM),
\ee
resembles the one appearing in
the theory of Bose gas with interaction, where the last term on the r.h.s.
describes interaction between particles and leads to appearing of quasiparticles
diagonalizing the Hamiltonian. Thus (\ref{HHH}) is subject to
another Bogoliubov transformation. Once operators $M^+$ and $M$ obey the Bose
statistics, then the general form of the transformation is expected to be
\be
\left\{
\begin{array}{lclcl}
m&=&uM&+&vM^+\\
m^+&=&uM^+&+&vM,
\end{array}
\right.
\label{simple1}
\ee
with $u$ and $v$ obeying the standard bosonic condition,
\be
u^2-v^2=1.
\label{simple2}
\ee

Of course, one needs to generalize transformation (\ref{simple1}), (\ref{simple2})
in order to take into account the nonlocality of the involved objects.
Such a generalization takes the form \cite{KNV}
\be
\begin{array}{c}
m^{+}_n(Q)=\ds\int\frac{\ds dq}{\ds 2\pi}\left\{M^+(q-Q,q)\vph_+^n(q,Q)+
M(q,q-Q)\vph_-^n(q,Q)\right\},\\
\\
m_n(Q)=\ds\int\frac{\ds dq}{\ds 2\pi}\left\{M(q-Q,q)\vph_+^n(q,Q)+
M^+(q,q-Q)\vph_-^n(q,Q)\right\},
\label{m}
\end{array}
\ee
\be
\begin{array}{c}
M(p-P,p)=\sum\limits_{n>0}\left\{m_n(P)\vph_+^n(p,P)-m^+_n(P)\vph_-^n(p,P)\right\},\\
\\
M^+(p-P,p)=\sum\limits_{n>0}\left\{m_n^+(P)\vph_+^n(p,P)-m_n(P)\vph_-^n(p,P)\right\},
\label{mM}
\end{array}
\ee
where $\vph_{\pm}$ stand for Bogoliubov-like functions $u$ and
$v$, so it is not surprise that they obey the following orthonormality and
completeness conditions, which are nothing but the generalization of
(\ref{simple2}):
\be
\begin{array}{rcl}
\ds\int\frac{\ds dp}{\ds 2\pi}\left(\vph_+^n(p,Q)\vph_+^{m}(p,Q)-\vph_-^n(p,Q)\vph_-^m(p,Q)
\right)&=&\delta_{nm},\\
&&\\
\ds\int\frac{dp}{2\pi}\left(\vph_+^n(p,Q)\vph_-^{m}(p,Q)-\vph_-^n(p,Q)\vph_+^m(p,Q)
\right)&=&0,
\label{norms}
\end{array}
\ee
\be
\begin{array}{rcl}
\ds\sum\limits_{n=0}^{\infty}\left(\vph^n_+(p,Q)\vph^n_+(k,Q)-\vph^n_-(p,Q)
\vph^n_-(k,Q)\right)&=&2\pi\delta\left( p-k\right),\\
&&\\
\ds\sum\limits_{n=0}^{\infty}\left(\vph^n_+(p,Q)\vph^n_-(k,Q)-\vph^n_-(p,Q)
\vph^n_+(k,Q)\right)&=&0.
\label{complet}
\end{array}
\ee

The functions $\vph_{\pm}^n$ play the role of the meson wave functions, moreover,
one can easily establish the physical meaning of both. Namely, $\vph_+^n$ describes the 
motion forward in time of the $q\bar q$ pair inside
meson, whereas $\vph_-^n$ is responsible for its motion backward in time.
We shall return to this issue later on when discussing the properties of the
bound-state equation.

The physical meaning of the transformation (\ref{m}) is quite obvious. Indeed, in the
theory with a nontrivial vacuum there are two ways to produce a quark-antiquark bound 
state. The first way, which works no matter if the vacuum is trivial or not, is to create 
the $q\bar q$ pair directly, by means of the operator $M^+\sim b^+d^+$. In the meantime, if
the vacuum is nontrivial and contains the chiral condensate $\langle \bar q q\rangle\neq 0$,
then one can \lq\lq borrow" a finite number of correlated quark-antiquark pairs from the
vacuum and to remove redundant particles, using the annihilation operator $M\sim db$. The
wave functions $\varphi_{\pm}$ describe the contributions of these two procedures. It
follows immediately from such a consideration that, {\it e.g.}, for the case of massive
quarks $\varphi_-$ should be small since the condensate of heavy quarks is suppressed by
the large quark mass.

\begin{figure}[t]
\centerline{\hspace*{1cm}\epsfig{file=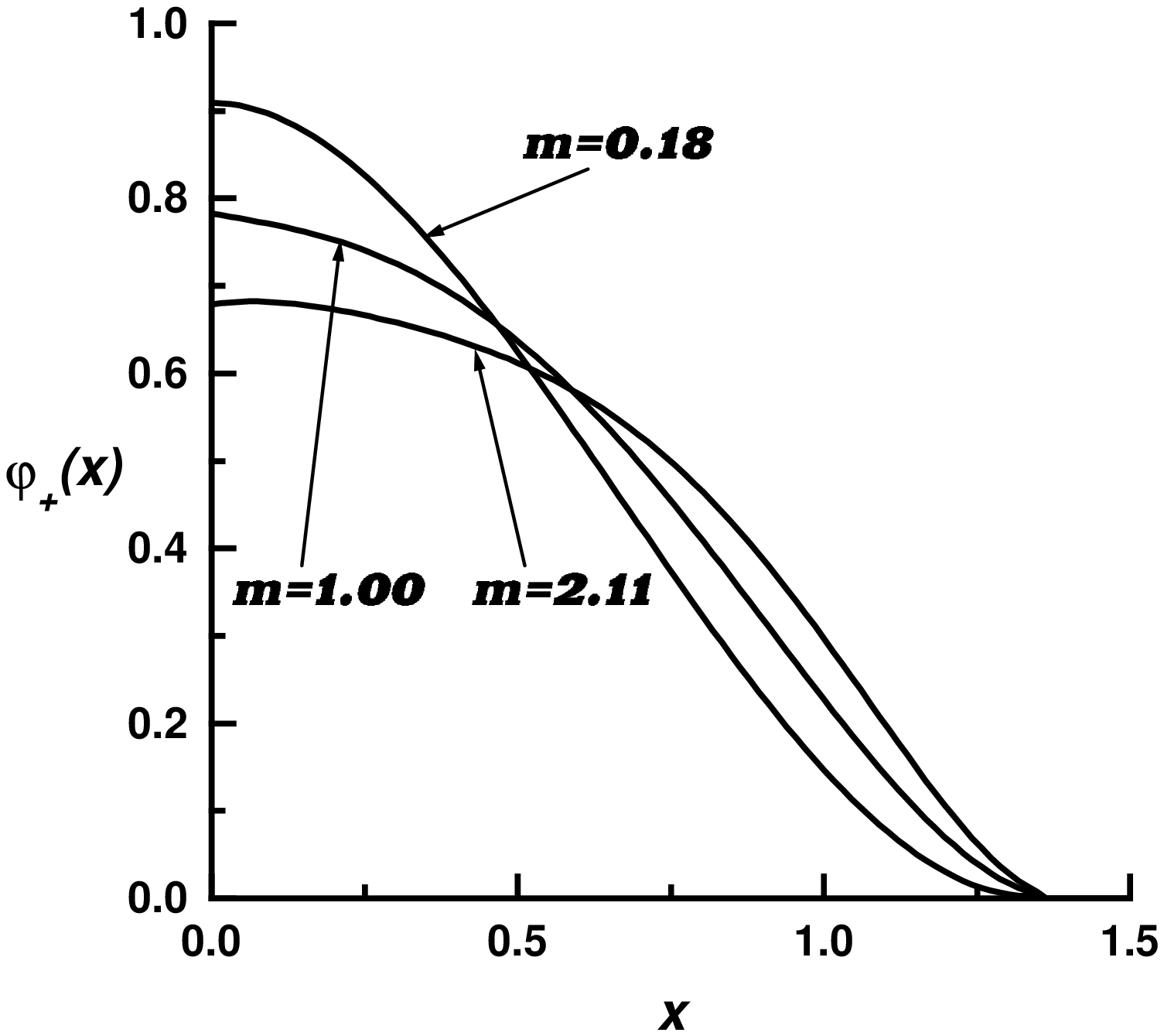,width=10cm}\hspace{-2.5cm}
            \epsfig{file=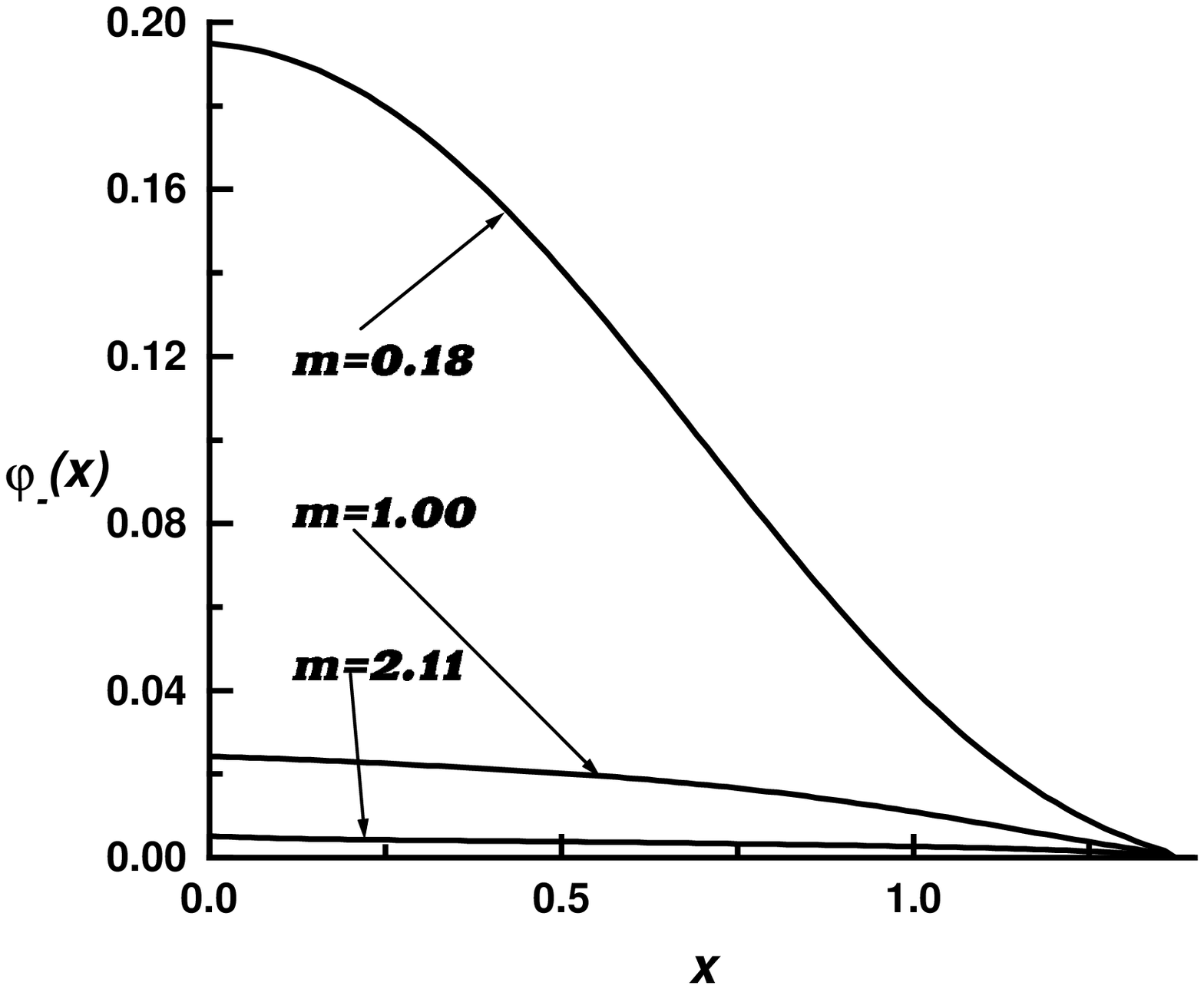,width=10cm}}
\caption{The $\vph_+$ and $\vph_-$ components of the ground-state wave function
in the meson rest frame for different
masses of the quark taken from \cite{Ming Li}. Note that $\vph_{\pm}$ are even
functions of $p$ for the ground state. The variable $x$ comes from the change 
$p=\tan(x)$, all dimensional quantities are given in the proper units of
$(2\gamma)^{1/2}$.}\label{figphi}
\end{figure}

It is easy to check that operators (\ref{m}) obey the standard 
bosonic commutation relations,
\be
\begin{array}{c}
\ds\left[m_n(Q)m^+_m(Q')\right]=2\pi\delta(Q-Q')\delta_{nm},\\
{}\\
\ds\left[\vphantom{m^+}m_n(Q)m_m(Q')\right]=\left[m^+_n(Q)m^+_m(Q')\right]=0,
\end{array}
\label{mcoms}
\ee
and diagonalize the Hamiltonian (\ref{HHH}),
\be
H=LN_C{\cal E}'_v +\sum\limits_{n=0}^{+\infty}\int\frac{dQ}{2\pi}Q^0_n(Q)m^+_n(Q)m_n(Q),
\label{HHHH}
\ee
if the wave functions $\vph_{\pm}^n$ are solutions to the bound-state equation
which we discuss in the next subsection. The vacuum energy in (\ref{HHHH})
contains extra contributions compared to that in (\ref{HHH}), besides that the vacuum
structure itself is changed, so that the real vacuum of the theory, $|\Omega\rangle$,
differs from $|0\rangle$ introduced in (\ref{bnd}) and they are related through a unitary
transformation. 

\subsection{The bound-state equation and properties of the mesonic wave functions}
\label{bseq}

As in case of the first Bogoliubov transformation performed in subsection
\ref{dressq}, when the gap equation (\ref{gap}) appeared as a condition of the
Hamiltonian diagonalization, the second, generalized, Bogoliubov transformation
described in the previous subsection also leads to an equation defining the
Bogoliubov-like functions $\vph_{\pm}^n$. This is nothing but the bound-state
equation for the mesonic spectrum of the model \cite{Bars&Green}\footnote{It is
straightforward to generalize the bound-state equation (\ref{BG}) for the
case
of a many-flavour theory. One needs to make the following modifications:
$$
E(p)\to E_{f_1}(p),\quad E(Q-p)\to E_{f_2}(Q-p),\quad \vph_{\pm}\to\vph_{\pm}^{f_1f_2},
$$
$$
C(p,k,Q)\to C_{f_1f_2}(p,k,Q)=\cos\frac{\ds\theta_{f_1}(p)-\theta_{f_1}(k)}{2}
\cos\frac{\theta_{f_2}(Q-p)-\theta_{f_2}(Q-k)}{2},
$$
$$
S(p,k,Q)\to S_{f_1f_2}(p,k,Q)=\sin\frac{\theta_{f_1}(p)-\theta_{f_1}(k)}{2}
\sin\frac{\theta_{f_2}(Q-p)-\theta_{f_2}(Q-k)}{2},
$$
where $f_1$ and $f_2$ stand for different quark flavours.\label{ftnt}}:
\be
\left\{
\begin{array}{l}
[E(p)+E(Q-p)-Q_0]\vph_+(Q,p)=\\
\hspace*{1cm}=\gamma\ds\vpint\frac{\ds dk}{\ds (p-k)^2}
\left[C(p,k,Q)\vph_+(Q,k)-S(p,k,Q)\vph_-(Q,k)\right]\\
{}\\

[E(p)+E(Q-p)+Q_0]\vph_-(Q,p)=\\
\hspace*{1cm}=\gamma\ds\vpint\frac{\ds dk}{\ds (p-k)^2}
\left[C(p,k,Q)\vph_-(Q,k)-S(p,k,Q)\vph_+(Q,k)\right].
\end{array}
\right.
\label{BG}
\ee

Unfortunately, analytical investigation of the system (\ref{BG}) is possible only
in some specific cases (we discuss one of them in the next subsection), so that
it is rather subject to numerical studies. In Fig.\ref{figphi}
taken from paper \cite{Ming Li} we
give numerical solutions for the ground state for different quark masses. 
It is clearly seen
that the $\vph_-$ component of the mesonic wave function is suppressed for large
masses of the quark in full agreement with the consideration presented 
above. 
It is also suppressed for highly excited states \cite{Ming Li}. 
Indeed, the case of higher excitations is very close to the quasiclassical
regime, where the answers for the spectra coincide with those of the quantum
mechanical problem of the two-dimensional string with quarks at the ends
\cite{2string}, which, in turn, is reduced to a quark potential model.

Let us demonstrate how the celebrated 't~Hooft equation comes out from (\ref{BG})
\cite{Bars&Green}.
In the above mentioned paper the Lorentz covariance was proved for the
colour-singlet sector of the theory, so that the spectrum of mass of the $q\bar q$ 
bound states, following from (\ref{BG}), should not change
when one performs boosts, even for the limiting case of the
boost into the infinite-momentum frame, $P\to\infty$. One can easily verify that
in this case $S(p,k,Q)\to 0$, whereas $C(p,k,Q)$ turns into
a step-like function, so that the region of integration on the r.h.s. of
(\ref{BG}) shrinks to a finite interval, $0<x<1$, with $x=p/P$ being the share of
the total momentum carried by the quark (the 't~Hooft variable). The $\vph_-$ component 
of the wave function dies out, and the system
(\ref{BG}) reduces to a single equation for $\vph_+$,
\be
M^2\vph(x)=\left(\frac{m^2-2\gamma}{x}+
\frac{m^2-2\gamma}{1-x}\right)\vph(x)-2\gamma\vpint\frac{dy}
{(y-x)^2}\vph(y),
\ee
\be
\vph(x)=\lim_{Q\to\infty}\sqrt{\frac{Q}{2\pi}}\vph_+(p,Q),
\label{vphx}
\ee
coinciding with the one derived by 't~Hooft in \cite{'tHooft}.

For further references we give here a couple of properties of the wave functions $\vph_{\pm}^n$:
\be
\vph_{\pm}^n(p\mp P,\mp P)=\vph_{\pm}^n(p,\pm P),
\label{prop1}
\ee
\be
\vph_{\pm}^n(P-p,P)=\eta_n\vph_{\pm}^n(p,P),
\label{prop2}
\ee
and the parity of the state is $(-1)^{\eta_n+1}$. The latter property
allows
one to 
classify all mesonic states into two groups: odd states, pions, with
$\eta_n=1$, and even states, $\sigma$'s, with $\eta_n=-1$. The odd and the even states
follow one by one in the spectrum starting from the lowest state, which is odd 
and it is
expected to be massless in the chiral limit. This state is nothing but the 
celebrated
chiral pion.
 
\subsection{Pionic solution}
\label{pn}
In this subsection we return to the bound-state equation (\ref{BG}) and find an
exact solution to it. Indeed, one can easily check that the set of the following 
two functions:
\be
\vph^{\pi}_{\pm}(p,Q) =N^{-1}_{\pi}\left(\cos\frac{\theta(Q-p)-\theta(p)}{2}\pm
\sin\frac{\theta(Q-p)+\theta(p)}{2}\right),
\label{pion}
\ee
\be
N^2_{\pi}(Q)=\frac{2}{\pi}Q
\label{pinorm}
\ee
satisfies the system (\ref{BG}) if the quark mass is put to zero. 
This solution turns out to be massless and thus it is nothing but the chiral pion ---
the lowest negative-parity state in the spectrum (see the discussion at the end of
the previous subsection). As one should anticipate, the norm of this state 
(\ref{pinorm}) vanishes in the pion rest frame, whereas in the infinite momentum
frame this solution reads
\be
\vph^{\pi}_-(p,Q)\mathop{\to}\limits_{Q\to\infty}0,\quad
\vph^{\pi}_+(p,Q)\mathop{\to}\limits_{Q\to\infty}\sqrt{\frac{2\pi}{Q}},\quad 0\leq
p\leq Q,
\ee
and, after an appropriate rescaling (see equation (\ref{vphx})), gives
$\varphi(x)=1$, {\it i.e.}, turns into the pionic solution found in
\cite{'tHooft,CCG}.

\subsection{The one-particle limit and the nonpotential quark dynamics}
\label{lnc}

Let us consider a
heavy-light system containing a static antiquark source placed at the origin
and a light quark. The most straightforward way to derive the bound-state
equation for this system is to take the one-body limit of the system (\ref{BG}). 
For the infinitely heavy flavour one has (see the footnote on the page \pageref{ftnt})
\be
E_{f_2}(p)=m_{f_2}\to\infty,\quad\cos\theta_{f_2}=1,\quad\sin\theta_{f_2}=0,\quad\vph_-^{f_1f_2}=0,
\ee
then the coefficients in (\ref{BG}) become:
\be
C_{f_1f_2}(p,k,Q)=\cos\frac{\ds\theta_{f_1}(p)-\theta_{f_1}(k)}{2},\quad
S_{f_1f_2}(p,k,Q)=0.
\ee

From now on, in this subsection, we shall suppress all flavour indices, having in mind
that the angle $\theta(p)$ and the dispersive law $E(p)$ are for the
light quark.

The system (\ref{BG}) reduces to a single equation \cite{KN1},
\be
\varepsilon\varphi(p)=E(p)\varphi(p)-
\gamma\vpint\frac{dk}{(p-k)^2}\cos\frac{\theta(p)-\theta(k)}{2}\varphi(k),
\label{alashroedinger}
\ee
where $\varphi=\vph_+$, $\varepsilon=Q_0-m_{f_2}$.

Note that the interaction in (\ref{alashroedinger}), given by the integral term
on the r.h.s., is essentially nonlocal and, moreover, there is no parameter (except
the mass of the quark) which
could allow one to treat this nonlocality perturbatively, expanding the cosine under
the integral. Meanwhile, if the quark is also heavy, then 
the cosine reduces to unity and 
equation (\ref{alashroedinger}) can be simplified even more, yielding the local
linearly rising potential between the constituents,
\be
\left(E_0(p)+\gamma |x|\right)\varphi(x)=\varepsilon\varphi(x),
\quad E_0(p)=\sqrt{p^2+m^2}\approx m+\frac{p^2}{2m}+\ldots.
\label{salpeter}
\ee

To get a deeper insight into the structure of the interaction in the 't~Hooft
model let us use another approach to the heavy-light system.

\subsection{A heavy-light system in the modified Fock-Schwinger gauge and 
the Lorentz nature of confinement}\label{qqq}

In this subsection we discuss another way to derive the one-particle limit
(\ref{alashroedinger}) of the bound-state equation (\ref{BG}), based on the
Schwinger-Dyson equation for QCD$_2$ in the modified Fock-Schwinger gauge 
(Balitsky gauge) \cite{KN1}. The advantage of this approach is a
possibility of its
generalization to QCD$_4$ if an appropriate model for the QCD vacuum is used
\cite{Sim2}.

First of all, note that the Coulomb gauge condition, $A_1(x_0,x)=0$, does not fix the
gauge completely allowing purely time-dependent gauge transformations. To
fix the
residual invariance we impose an extra condition, 
$A_0(x_0,x=0)=0$,\footnote{In 3+1 this gauge is usually introduced via conditions
$A^a_0(x_0,\vec{0})=0$ and $\vec{x}\vec{A}^a(x_0,\vec{x})=0$ \cite{Balgauge}.}
which obviously breaks translational invariance, but turns out very convenient if an 
infinitely heavy (static) particle is involved. In case of the system 
containing a light quark and a static antiquark source the origin can be associated
with the latter, which appears to play an extremely passive role in the
interaction providing the white colour of the entire object. 
The Green's function of the static antiquark is independent of the gluonic field in
this gauge, 
$S_{\bar Q}(x,y|A)=S_{\bar Q}(x-y)$, and takes the form:
\be
S_{\bar Q}(x)= \hat{1}(-i)\left(\frac{1+\gamma_0}{2} \theta (-x_0)
e^{iMx_0}+\frac{1-\gamma_0}{2}\theta(x_0)e^{-iMx_0}\right)\delta(x),
\label{SaQ}
\ee
where $\hat{1}$ denotes unity in the colour space. The Green's function 
(\ref{SaQ}) contains also an
infinite set of corrections of the form $\frac{1}{M^n}\frac{d^n}{dx^n}\delta(x-y)$ which
die out in the limit $M\to\infty$.

The Green's function of the colourless $q\bar Q$ system can be written in the
following form:
\be
S_{q\bar Q}(x,y)=\frac{1}{N_C}\int D\psi D{\bar\psi}DA_{\mu}
\exp{\left\{-\frac{i}{4}\int d^2xF_{\mu\nu}^{a2}+i\int d^2x
{\bar\psi}(i\hat \partial -m -\hat A)\psi \right\}}\times
\label{SqQ}
\ee
$$
\times{\bar\psi}(x) S_{\bar Q} (x-y)\psi(y),
$$
so that it turns out possible to integrate out the gluonic field arriving at
a Schwinger-Dyson equation for the light-quark Green's function $S(x,y)$,
\be
(i\hat{\partial}_x-m)S(x,y)-(2\pi)^2\gamma\int d^2z
\gamma_0S(x,z)\gamma_0{\tilde D}_{00}(x,z)S(z,y)=\delta^{(2)}(x-y).
\label{SD}
\ee

Note that $S(x,y)=\frac{1}{N_C}S_{\alpha}^{\alpha}(x,y)$ possesses all
properties of the full $q\bar Q$ Green's function due to the passive role of the antiquark
discussed above. Then both, one-particle ({\it e.g.}, the chiral condensate), as well as
two-particle ({\it e.g.}, the spectrum of bound states) properties of the system
can be extracted from the single function $S(x,y)$. A special attention is to be payed to the gluonic
propagator ${\tilde D}_{00}(x,y)$, which looks similarly to that in the Coulomb gauge (\ref{D}) but
contains extra terms breaking the translational invariance and encoding the light-quark
interaction with the static antiquark,
\be
{\tilde D}_{00}(x_0-y_0,x,y)=-\frac{i}{2}(|x-y|-|x|-|y|)\delta(x_0-y_0)
\equiv K(x,y)\delta(x_0-y_0),
\label{K}
\ee
or in the momentum space,
$$
K(p,q)=K^{(1)}(p,q)+K^{(2)}(p,q),
$$
\be
K^{(1)}(p,q)=\frac{i}{p^2}\delta(p-q),
\label{K1}
\ee
\be
K^{(2)}(p,q)=-\frac{i}{q^2}\delta(p)-\frac{i}{p^2}\delta(q),
\label{K2}
\ee
where we have separated the local and the nonlocal parts.

From now on two different strategies can be adopted, which finally lead to the same
equation for the spectrum of the heavy-light system. The first approach is based on a
diagrammatic technique with two different internal lines prescribed to the local and the
nonlocal parts of the kernel \cite{KN1}. It turns out that, in spite of the nonlinearity
of the equation (\ref{SD}), the two parts of the kernel can be considered separately
due to very peculiar properties of rainbow diagrams \cite{KN1}. Thus the local part $K^{(1)}$
defines the mass operator $\Sigma(p)$, which can be naturally parametrized by means of the 
dressed quark dispersive law $E(p)$ and the chiral angle $\theta(p)$,
\be
\Sigma(p)=\left[E(p)\cos\theta(p)-m\right]+\gamma_1\left[E(p)\sin\theta(p)-p\right],
\label{Sigma}
\ee
with the system of coupled equations (\ref{system}) being the selfconsistency condition of
such a parametrization. Then the nonlocal part $K^{(2)}$ eventually gives the 
bound-state equation (see \cite{KN1} for the details). 

Here we choose the other strategy based on the Foldy--Wouthoysen transformation of
equation (\ref{SD}) \cite{ConfIII}. First, we rewrite this equation in the momentum space
and use a spectral decomposition for the light-quark Green's function,
\be
S(q_{10},q_1,q_{20},q_2)=2\pi\delta(q_{10}-q_{20})\left(
\sum_{\varepsilon_n>0}\frac{\varphi_n^{(+)}(q_1)\bar{\varphi}_n^{(+)}(q_2)}
{q_{10}-\varepsilon_n+i0}
+\sum_{\varepsilon_n<0}\frac{\varphi_n^{(-)}(q_1)\bar{\varphi}_n^{(-)}(q_2)}
{q_{10}+\varepsilon_n-i0}\right),
\label{specS}
\ee
where the positive- and the negative-energy solutions $\varphi_n^{(\pm)}$ have been introduced.
To proceed further we assume that a Foldy-Wouthoysen operator, 
$T_F(p)=e^{-\frac12\theta_F(p)\gamma_1}$,
diagonalizing equation (\ref{SD}), exists and that the angle $\theta_F$ is the same for all
$n$'s. With such an assumption one has
\be
\varphi_n^{(+)}(p)=\varphi_n^0(p)T_F(p){1\choose 0},\quad
\varphi_n^{(-)}(p)=\varphi_n^0(p)T_F(p){0\choose 1},
\label{def0}
\ee
\be
\int\frac{dp}{2\pi}\vph_n^0(p)\vph_m^0(p)=\delta_{nm},\quad 
\sum_n\vph_n^0(p)\vph_n^0(q)=2\pi\delta(p-q),
\label{simcom}
\ee
so that the following relation holds true for the Green's function (\ref{specS}):
\be
\int\frac{d\omega}{2\pi}S(\omega,q_1,q_2)=-i\pi\delta(q_1-q_2)[
\cos\theta_F(q_1)-\gamma_1\sin\theta_F(q_1)],
\ee
where $\omega=q_{10}-q_{20}$.

The Schwinger-Dyson equation (\ref{SD}) reduces then to a
Dirac-type equation in the Hamiltonian form,
\be
(\alpha p +\beta m)\varphi_n^0(p)-\frac{i\gamma}{2}
\int dqdk(\beta \cos\theta_F(q)+\alpha\sin\theta_F(q))K(p-q,k-q)
\varphi_n^0(k)=\varepsilon_n\varphi_n^0.
\label{beforeF}
\ee

The local part of the interaction in (\ref{beforeF}), generated by $K^{(1)}$, leads to dressing
of the light quark described by the Bogoliubov-Valatin angle $\theta(p)$ and the dressed
quark
dispersive law $E(p)$ obeying the system (\ref{system}). Therefore, one comes to the
conclusion that the Foldy angle $\theta_F(p)$ coincides with the Bogoliubov-Valatin one,
\be
\theta_F(p)=\theta(p).
\label{coin}
\ee

The nonlocal part of the interaction in (\ref{beforeF}), which stems 
from $K^{(2)}$, is also
diagonalized then, so that one ends with a Schr{\"o}dinger-type equation,
\be
\varepsilon_n\varphi^0_n(p)=E(p)\varphi^0_n(p)-
\gamma\vpint\frac{dk}{(p-k)^2}\cos\frac{\theta(p)-\theta(k)}{2}\varphi^0_n(k),
\label{alash2}
\ee
which coincides with (\ref{alashroedinger}). 

Comparing bound-state equation (\ref{alash2}) with (\ref{BG}) one finds that 
$\varphi^0$ plays the role of the $\varphi_+$ component of the heavy-light system 
wave functions, whereas $\varphi_-$ vanishes due to presence of the
infinitely massive antiquark. Thus relations (\ref{simcom}) follow immediately from
(\ref{norms}) and (\ref{complet}) with all $\varphi_-$'s put to zero. 

It is instructive to note that in the Coulomb (as well as Balitsky) gauge the 't~Hoof model is 
totally defined by only one nontrivial 
function $\theta(p)$, solution to the gap equation (\ref{gap}), which plays a
threefold role: 
\begin{itemize}
\item it defines the Bogoliubov-Valatin transformation from bare to
dressed quarks; 
\item it gives the Foldy angle, which diagonalizes the interquark 
interaction in the model; 
\item it entirely defines all quantities in the model, including the
bound-state equation.
\end{itemize}

Several comments concerning equation (\ref{SD}) are in order here. The first one
deals with the generalization of (\ref{SD}) to the four-dimensional case. The
attentive reader may notice that the only two-dimensional constituent of 
(\ref{SD}) is the gluonic propagator ${\tilde D}_{\mu\nu}(x,z)$ taken in the form 
(\ref{K}). The
equation itself survives in the case of QCD$_4$, if one has an appropriate form of the
bilocal gluonic correlator ${\tilde D}_{\mu\nu}(x,z)$ in the given gauge, 
and  
some arguments exist, why 
higher orders correlators, which lead to many-fermion vertices higher than four, 
can be neglected (see \cite{Casimir} for details). 

Another interesting issue concerning
equation (\ref{SD}) is that one cannot simplify the interaction kernel
substituting
\be
\gamma_0S\gamma_0\to\gamma_0S_0\gamma_0,
\label{SS0}
\ee
as proposed in \cite{Nora}. The reason for this failure is discussed in detail 
in \cite{PotReg}
and comes from the fact that the real parameter defining the substitution
(\ref{SS0}), with the consequent expanding of $S_0$ in powers of the one-dimensional
momentum, is the product of the quark mass and the gluonic correlation length.
The latter parameter defines also the radius of the string formed between the
colour constituents in the theory. A simple dimensional analysis demonstrates
that strings are infinitely thin in 1+1, as the system has too low dimension to
allow them to swell. One can arrive at the same conclusion inspecting the
two-dimensional correlator 
$\langle FF\rangle\equiv Tr\langle F_{\mu\nu}(x)\Phi(x,y)F_{\rho\sigma}(y)\Phi(y,x)
\rangle$, where $\Phi(x,y)$ is the standard parallel transporter along an arbitrary
path between the points $x$ and $y$, which provides the gauge invariance of the entire
nonlocal object \cite{cumul}
(see also the review paper \cite{YuArev} for the detailed description of the formalism).
Using the gluonic propagator (\ref{D}), one easily finds that $\langle FF\rangle$ is
proportional to the two-dimensional $\delta$-function in the configuration space,
\be
\langle FF\rangle\sim\delta^{(2)}(x-y),
\ee
{\it i.e.}, it has zero correlation length $T_g$. Thus the product $mT_g$ is
identically zero in two-dimensional QCD, which makes the interaction
essentially nonlocal and the quark dynamics becomes not potential
\cite{PotReg,ConfIII}. 

Finally, equation (\ref{beforeF}) answers, at least in QCD$_2$, 
the long-standing question on the Lorentz nature of confinement. One should distinguish
between the Lorentz structure of the confining interaction which is of the
$\gamma_0\times\gamma_0$ type (see (\ref{SD})) and the {\it effective} interaction, which enters the
Dirac-like
equation (\ref{beforeF}), describing the bound-state problem. As clearly seen from (\ref{beforeF}), 
the latter contains only effective scalar (terms $\sim\gamma_0$) and space vector (terms
$\sim\gamma_0\gamma_1=\gamma_5$) interactions.

\subsection{The chiral properties of the model in the Hamiltonian approach}
\label{cpm}

In this subsection we discuss the chiral properties of the model which are
highly nontrivial and in many features resemble those of four-dimensional
QCD. 

Most studies of the 't~Hooft model have been performed in the light-cone gauge,
$A_-=0$, which leads to a perturbative vacuum and to a simpler bound-state
equation. In the meantime, one has to employ rather sophisticated methods to
discuss the chiral limit of the model, when all functions and distributions
become extremely singular, and the whole range of the 't~Hooft variable
definition, $0<x<1$, is squeezed to small intervals near the boundary points
$x=0$ and $x=1$, which deliver all nontrivial content of the theory. 
The Hamiltonian approach in the Coulomb gauge 
developed above is free of this drawback.
Indeed,
to calculate a matrix element of any operator between mesonic states, the only relevant
ones in the weak regime of the model, one is to rewrite the 
above-mentioned operator in terms of operators $m^+$ and $m$ introduced in
(\ref{m}) and to use the second quantization technique to evaluate 
directly the
matrix element. The result is always expressed in terms of trigonometric
functions of the angle $\theta(p)$ and integrals of them, which can be worked
out analytically in some cases, or treated numerically. Anyway, with the numerical solution for $\theta$ found in
\cite{Ming Li} any value in the model appears calculable. 

Let us make just one more comment
concerning the vacuum of the theory. As it was mentioned above, the
true vacuum of the model is the mesonic one annihilated by the mesonic
operators,
\be
m_n(P)|\Omega\rangle=0,
\ee
for any $n$ and any total momentum $P$. By a unitary transformation this state is related
to the quark vacuum defined by relations (\ref{bnd}),
\be
|\Omega\rangle =U|0\rangle.
\ee

Despite of the fact that the explicit form of the operator $U$ is unknown,
the difference between the averages calculated with the help of the mesonic and
the quark vacua turns out to be suppressed in the large-$N_C$ limit, 
so that it plays the role of a small correction and, hence, lies beyond
the scope of the present paper. Thus for practical calculations one is free to use any 
of the above two vacua.

\subsubsection{The chiral condensate}

A crucial test for the chiral symmetry, to see if it is spontaneously broken or
not, is the chiral condensate $\langle {\bar q}q\rangle$. If this average
does not vanish for the vacuum state, then the whole tower of physical states 
will lack the chiral symmetry respected by the Hamiltonian of the theory,
so that this symmetry appears spontaneously broken. For the simplest case of only one
quark flavour the chiral symmetry breaking reads $U(1)_L\times U(1)_R\to U(1)_V$, 
and it is very important that the $U(1)_A$ invariance is broken spontaneously in the
't~Hooft model, in contrast to QCD$_4$ where this breaking is explicit due to the axial
anomaly. Indeed, the two-dimensional  anomaly is proportional to
the colour trace of the coloured object $\tilde{F}\sim \varepsilon_{\mu\nu}F_{\mu\nu}$,
which obviously vanishes (in the meantime, the axial anomaly does exist in the
two-dimensional QED, known as the Schwinger model, where no colour trace should
be taken).

Now we are in the position to evaluate the chiral condensate for the 't~Hooft
model. Following the general approach described at the beginning of this
subsection, one can find
$$
\langle\bar{q}q\rangle=
\left\langle\Omega\left|\bar q_{\alpha}(x)q^{\alpha}(x)\right|\Omega\right\rangle\tooo
\left\langle 0\left|\bar q_{\alpha}(x)q^{\alpha}(x)\right|0\right\rangle=\hspace*{2cm}
$$
\be
\hspace*{2cm}=N_C\int\frac{dk}{2\pi}Tr\left\{\gamma_0\Lambda_{-}(k)\right\}=
-\frac{N_C}{\pi}\int\limits_{0}^{+\infty}dk\cos\theta(k).
\label{condensate}
\ee

It is instructive to arrive at the same result using another approach --- namely,
the definition of the condensate via the light-quark Green's function. In spite of the
fact that the single quark Green's function is a gauge variant object and,
hence, it is not physical by itself, when taken with the coinciding arguments and
summed up over the colours, it readily gives the chiral condensate\footnote{A
special care should be taken at the both stages. Indeed, the Green's function
contains a discontinuity at $x=y$, so that one should approach this limit
either from
the side of larger, or smaller $y$'s. On the other hand, if the condensate is
calculated beyond the chiral limit, then the logarithmically-ultraviolet-divergent 
perturbative contribution,
proportional to the quark mass, should be subtracted from (\ref{condensate_1}).}
\be
\langle{\bar q}q\rangle=-i\trr S_{\alpha}^{\alpha}(x,y).
\label{condensate_1}
\ee

Once the relation (\ref{condensate_1}) is gauge invariant, then let us choose the
modified Fock-Schwing\-er gauge discussed above. The quark Green's function,
$S(x,y)=\frac{1}{N_C}S_{\alpha}^{\alpha}(x,y)$, 
is the solution to equation (\ref{SD}) \cite{KN1}. Using its spectral
decomposition in the coordinate space,
$$
S(x_0-y_0,x,y)=-i\sum_n\psi_{n}^{(+)}(x)\bar{\psi}_{n}^{(+)}(y)
e^{-i\varepsilon_n(x_0-y_0)}\theta(x_0-y_0)+
$$
\be
+i\sum_n\psi_{n}^{(-)}(x)\bar{\psi}_{n}^{(-)}(y)
e^{i\varepsilon_n(x_0-y_0)}\theta(y_0-x_0),
\label{101}
\ee
\be
\psi_{n}^{(\pm)}(x)=\int_{-\infty}^{\infty} 
\frac{dp}{2\pi}\;\varphi_{n}^{(\pm)}(p)
e^{ipx},
\ee
where $\varepsilon_n$ are the eigenenergies defined by equation (\ref{alash2}), 
substituting the Foldy-rotated wave functions (\ref{def0}),
\be
\begin{array}{l}
\varphi_n^{(+)}(p)=T^+(p){\tilde \varphi}_n^{(+)}(p)=
\varphi_n^0(p)T^+(p){1\choose 0},\\
{}\\
\varphi_n^{(-)}(p)=T^+(p){\tilde \varphi}_n^{(-)}(p)=
\varphi_n^0(p)
T^+(p){0\choose 1},
\end{array}
\label{102}
\ee
and, finally, using the simplified completeness 
condition (\ref{simcom}) for the set $\{\varphi_n^0(p)\}$,
one reproduces the result (\ref{condensate}) \cite{KN1}.

It is easily seen from the definition (\ref{condensate_1}) that diagrammatically the
chiral condensate can be represented as a closed fermion line which begins and ends at
the coinciding points. Such way both, the motion forward in time 
(the positive-energy solutions) and the one backward in time 
(the negative-energy solutions), are equally
important for the condensate. At first glance this statement contradicts the observation
made above, that all
$\varphi_-$ components vanish for the heavy-light system. Solution of this problem can be
found in properties of the bound-state equation (\ref{BG}). Indeed, there
are, in fact, two sets of solutions to the system (\ref{BG}), 
with $Q_0>0$ and $Q_0<0$, trivially connected with one another,
\be
\varphi_+^{-n}(p,P)=\varphi_-^{n}(p,P),\quad \varphi_-^{-n}(p,P)=\varphi_+^{n}(p,P),
\ee
where positive $n$'s numerate states with $Q_0>0$, and negative $n$'s are prescribed to the
states with $Q_0<0$. Therefore, these are $\varphi_+^{-n}(p,P)=\varphi_-^{n}(p,P)$ $(n>0)$ to
vanish for the heavy-light system, whereas the two remaining wave functions 
$\varphi_-^{-n}(p,P)=\varphi_+^{n}(p,P)$ describe the propagation of the $q\bar Q$ 
system either forward or backward in time without Zitterbewegung. They both contribute 
on equal footing to the chiral condensate (\ref{condensate_1}).

From (\ref{condensate}) one can see that the properties of the solution 
for $\theta$ to the gap
equation (\ref{gap}) are of paramount importance for the chiral symmetry
breaking. If the integral on the r.h.s. in (\ref{condensate}) vanishes, then we
are in the phase of the theory with the restored chiral symmetry. 

It was mentioned in subsection \ref{dressq} that there exist two different 
solutions to the gap equation (\ref{gap}) in the chiral limit. One of them, found
analytically in \cite{Bars&Green}, gives $|\theta(p)|=\pi/2$ and, hence,
$\cos\theta(p)=0$ everywhere, so that the chiral condensate (\ref{condensate}) 
vanishes for this solution (as was discussed above, this phase has an infinite energy and,
hence, never realises). Luckily it is
not so for the numerical solution found in \cite{Ming Li} and depicted in 
Fig.\ref{figtheta}. Substituting it into (\ref{condensate}) and working
out the integral numerically one finds:
\be
\langle\bar{q}q\rangle_{m=0}=-0.29N_C\sqrt{2\gamma},
\label{condensate_2}
\ee
that coincides with the results found in \cite{Zhitnitskii}\label{exactcon}\footnote{The
corresponding result from \cite{Zhitnitskii} reads
$$
\langle\bar{q}q\rangle_{m=0}=-\frac{N_C\sqrt{\gamma}}{\sqrt{6}},
$$
that numerically coincides with (\ref{condensate_2}), thus giving evidence 
that various integrals of $\theta$ can be found not only numerically, but also
in the form  of irrational numbers.}. The chiral
symmetry is spontaneously broken in this phase of the theory and the pion, found
in the subsection \ref{pn}, is, indeed, the corresponding Goldstone boson.
Note that the chiral condensate for the 't~Hooft model is known analytically 
for any value of the quark mass \cite{Burkardt}.

Now we can return to formula (\ref{EA2}) for the vacuum energy and to rewrite it using a
more physically transparent language. Indeed, applying the same transformation, 
$\theta(p)\to\theta(p/A)$, to the chiral condensate (\ref{condensate}), one easily finds
that $\Sigma\equiv \langle \bar q q\rangle$ scales linearly with $A$,
\be
\Sigma \to A \Sigma,
\ee 
so that the mute parameter $A$ can be changed for the chiral condensate and 
relation (\ref{EA2}) can be written as
\be
\Delta{\cal E}_v=C_1'\left[\frac12\left(\frac{\Sigma_{\phantom 0}}{\Sigma_0}\right)^2-
\ln\left|\frac{\Sigma_{\phantom 0}}{\Sigma_0}\right| \right]+\gamma C_3',
\label{ve2}
\ee
where the minimum of the vacuum energy is reached for $\Sigma=\Sigma_0$ given by (\ref{condensate_2}).
If the quark mass does not vanish, then the vacuum energy density (\ref{ve2}) acquires an extra
contribution, 
\be
\Delta{\cal E}_v\to \Delta{\cal E}_v+m\left(\frac{\Sigma_0}{N_C}\right)
\left(\frac{\Sigma_{\phantom 0}}{\Sigma_0}\right),
\label{addtrm}
\ee
which explicitly breaks the invariance of $\Delta{\cal E}_v$ with respect to the change
$\Sigma\to -\Sigma$. Then the lowest (pionic) excitation over the vacuum with the wrong 
sign of the 
condensate acquires an imaginary mass and becomes the tachyon as it follows immediately
from the Gell-Mann-Oakes-Renner relation (see equation (\ref{glm})).

Minimization procedure for the
vacuum energy in presence of the mass term (\ref{addtrm}) leads to a more complicated equation for 
$\Sigma$, so that $\Sigma_0$ does not provide the minimum anymore. 

\subsubsection{The pion decay constant}

In this subsection we derive the decay constant $f_{\pi}$ for the chiral pion. Using the
standard definition for it,
\be
\left.\left\langle\Omega\right.\right|J_{\mu}^5(x)\left|\left.\pi(Q)\right.
\right\rangle=f_{\pi}Q_{\mu}\frac{e^{-iQx}}{\sqrt{2Q_0}},
\label{fpi}
\ee
where
\be
J_{\mu}^5(x)=\bar{q}(x)\gamma_{\mu}\gamma_5q(x),
\label{J5def}
\ee
one can calculate the matrix element explicitly using the technique described
above, so that the result reads
\be
f_{\pi}=\sqrt{\frac{N_C}{\pi}}.
\label{ffpi}
\ee

Note that the pion decay constant (\ref{ffpi}) is dimensionless in the 't~Hooft model, which
drastically differs in this point from QCD$_4$, where this constant is
dimensional and it appears rather small at the hadronic scale ($93MeV$). 
Thus in the four-dimensional case the pion decay constant
defines a new scale for the effective low-energy QCD --- the chiral perturbation
theory, which cannot be developed in the 't~Hooft model in
view of the dimensionlessness of $f_{\pi}$.

\subsubsection{Partial conservation of the axial-vector current (PCAC)}

In this subsection we derive explicitly the PCAC relation for the 't~Hooft
model.

Starting from the definition of the axial-vector current (\ref{J5def}),
we use representation (\ref{quark_field}) for the quark fields and introduce 
bosonic operators (\ref{operators}) after an appropriate ordering of the antiquark
creation and annihilation operators $d^+$ and $d$. Leaving only the leading terms in the
$1/N_C$ expansion, one arrives at
\be
J_{\mu}^5(x)=N_C\int\frac{dk}{2\pi}v^+(-k)\gamma_{\mu}\gamma_5v(-k)+\hspace*{5cm}
\ee
$$
\sqrt{N_C}\int\frac{dpdP}{(2\pi)^2}e^{-iPx}\left[M^+(p,p-P)u^+(p-P)
\gamma_{\mu}\gamma_5v(-p)+M(p-P,p)v^+(P-p)\gamma_{\mu}\gamma_5u(p)\right].
$$

The explicit form of the quark amplitudes $u$ and $v$ given by (\ref{unv})
together with the definition of the mesonic creation and annihilation operators
(\ref{m}) allows one to proceed further and to rewrite the components of the
axial-vector current in the form
\be
J_0^5(x)=2\sqrt{N_C}\int\frac{dP}{2\pi}N_{\pi}e^{-iPx}\sum_{n=0}^{\infty}(m_n(P)-m^+_n(-P))
\int\frac{dp}{2\pi}g_{\pi}(p,P)f_n(p,P),
\ee
\be
J_1^5(x)=2\sqrt{N_C}\int\frac{dP}{2\pi}N_{\pi}e^{-iPx}\sum_{n=0}^{\infty}(m_n(P)+m^+_n(-P))
\int\frac{dp}{2\pi}f_{\pi}(p,P)g_n(p,P),
\ee
where
\be
f_n(p,P)=\frac12(\vph^n_+(p,P)-\vph^n_-(p,P)),\quad
g_n(p,P)=\frac12(\vph^n_+(p,P)+\vph^n_-(p,P)).
\label{gfdef}
\ee

From (\ref{norms}) one easily finds that
\be
\int\frac{dp}{2\pi}g_{\pi}(p,P)f_n(p,P)=\int\frac{dp}{2\pi}f_{\pi}(p,P)g_n(p,P)=
\frac{1}{4}\delta_{n\pi},
\ee
{\it i.e.}, in the chiral limit the axial-vector current couples only to pions,
and it can be written as
\be
J_{\mu}^5(x)=i\sqrt{\frac{N_C}{\pi}}\partial_{\mu}\int\frac{dP}{2\pi}\frac{1}{\sqrt{2P_0}}
\left(e^{-iPx}m_{\pi}(P)+e^{iPx}m_{\pi}^+(P)\right)=if_{\pi}\partial_{\mu}
\Psi_{\pi}(x),
\label{pcac}
\ee
where the second-quantized wave function of the pion in the coordinate space 
$\Psi_{\pi}(x)$ is introduced. 

Relation (\ref{pcac}) is nothing but the celebrated partial conservation of the
axial-vector current (PCAC), whose operator form is usually formulated as
a 
hypothesis in QCD$_4$. In the 't~Hooft model the latter can be proved
explicitly and the Hamiltonian approach to the model turns out the most natural
environment for this task. 

It is instructive to note that the form of the pionic solution can be easily guessed
even before the $:H_4:$ part of the Hamiltonian is
taken into account. Indeed, if the chiral symmetry is spontaneously broken, 
then the corresponding charge does not commute with the Hamiltonian,
\be
[Q_5H]\neq 0,\quad Q_5=\int dx J_0^5(x),
\ee
so that, if the Hamiltonian is diagonalized in the quark sector, then 
$Q_5$ contains an anomalous term,
\be
Q_5\sim\int\frac{dp}{2\pi}[b^+(p)d^+(-p)+d(-p)b(p)]\cos\theta(p),
\ee
with the coefficient $\cos\theta(p)$ being right the pion wave function 
in the rest frame
(see equations (\ref{pion}) and (\ref{pisol2})).

\section{Matrix approach}

In spite of evident technical advantages and physical transparency of the 
Hamiltonian approach developed and
exploited in the previous section, it has a number of disadvantages. Among those
are a rather tedious algebra and not straightforward connection to the
diagrammatic technique which is very convenient in studies of variety of
hadronic
processes. In the present section we develop a matrix approach to the
't~Hooft model, which allows one to simplify considerably investigations of some
phenomena, {\it e.g.}, this technique will be effectively used in the next section, where the
Ward identities and the strong hadronic decays are discussed.

The section is organized as follows. In subsection 3.1, following \cite{Bars&Green}, 
we introduce the matrix wave function and derive the bound-state equation for it. In 
subsection 3.2 we study properties of the matrix Hamiltonian --- namely, its Hermiticity
and the Hilbert space of its definition. Chiral properties of the 't~Hooft model are the
subject of the next subsection 3.3. We establish the Gell-Mann-Oakes-Renner relation,
discuss the pionic wave function beyond the chiral limit, and return to the calculation of
the pion decay constant. In the consequent subsections 3.4, 3.5, and 3.6 we derive the
quark-quark scattering amplitude and Ward identities for the vector and axial-vector
currents and find the pionic vertex, respectively.

\subsection{Matrix wave functions and matrix bound state equation}

In this subsection we briefly recall the method and the results of the paper 
\cite{Bars&Green} based on the diagrammatic approach to the theory.

\begin{figure}[t]
\begin{center}
\epsfxsize=16cm
\epsfbox{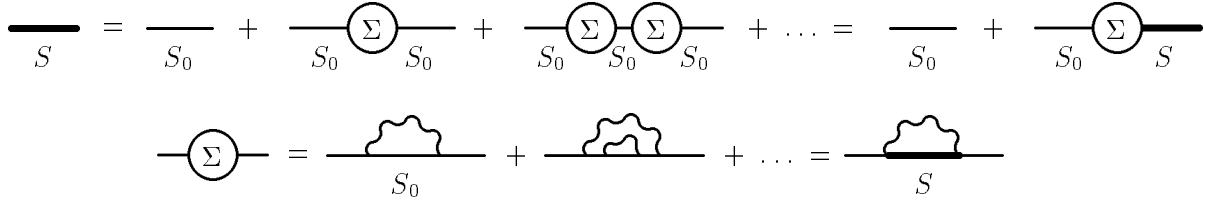}
\caption{Graphical representation of the equations for the dressed quark propagator and
the quark mass operator.}\label{figsig}
\end{center}
\end{figure}

At the first step we define the mass operator $\Sigma$ as a sum of planar
diagrams (see Fig.\ref{figsig}) which contribute to the dressed quark
propagator $S(p_{\mu})$,\footnote{We use the argument $p_{\mu}$ as a shorthand notation for
$(p_0,p)$. If not stated explicitly, then dependence only on the one-dimensional spatial
momentum is meant, like in $\Sigma (p)$.}
\be
S(p_{\mu})=\frac{1}{\hat{p}-m-\Sigma(p)+i\varepsilon},
\label{S}
\ee
\be
\Sigma(p)=\frac{i\gamma}{2\pi}\int\frac{dk_0dk}{(p-k)^2}\gamma_0S(k_{\mu})\gamma_0.
\label{Sigma2}
\ee

Note that due to the instantaneous type of the interaction the integration 
over $k_0$ is trivial and the mass operator depends only on the spatial
component of the momentum. Using the same parametrization as in 
(\ref{Sigma}) one immediately arrives at the gap equation in the form (\ref{gap})
and the definition of $E(p)$ via $\theta(p)$ coinciding with (\ref{E}).

\begin{figure}[t]
\begin{center}
\epsfxsize=14cm
\epsfbox{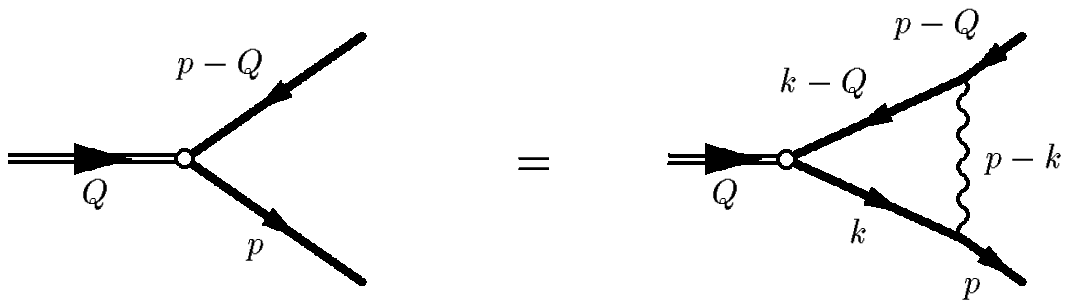}
\caption{Graphical representation for the Bethe-Salpeter equation (\ref{BSeq}).}\label{figbs}  
\end{center}
\end{figure}

As the second step a homogeneous Bethe-Salpeter equation is used, which is
diagrammatically represented in Fig.\ref{figbs} and defines the spectrum of the
quark-antiquark
bound-states. The fat lines denote the dressed quark propagators
(\ref{S}) whereas the meson-quark-antiquark vertices are described by the function
$\Gamma(p,P)$. It is also convenient to introduce a 
modified vertex $\tilde{\Gamma}(p_{\mu},P_{\mu})$,

\be
\tilde{\Gamma}(p_{\mu},P_{\mu})=-i\gamma S(p)\Gamma(p,P) S(p-P),
\label{tGamma}
\ee
and a matrix
wave function $\Phi(p,Q)$ defined in the standard way \cite{Bars&Green},
\be
\Phi(p,Q)=\int\frac{dp_0}{2\pi}\tilde{\Gamma}(p_{\mu},Q_{\mu})
=\int\frac{dp_0}{2\pi}\tilde{\Gamma}(p_0-Q_0,p,Q).
\label{Phidef}
\ee

The equation corresponding to the diagrams in Fig.\ref{figbs} reads
\be
\tilde{\Gamma}(p_{\mu},Q_{\mu})=\frac{i\gamma}{2\pi}\int\frac{dk_0dk}{(p-k)^2}S(p_{\mu})
\gamma_0\tilde{\Gamma}(k_{\mu},Q_{\mu})\gamma_0S(p_{\mu}-Q_{\mu}),
\label{BSeq}
\ee
or, after integrating both sides of this equation over $p_0$, introduction of the
wave function $\Phi$ according to relation (\ref{Phidef}), and performing
simple algebraic transformations, one arrives at the bound-state equation in the matrix
form,
$$
Q_0\Phi(p,Q)=(\gamma_5p+\gamma_0m)\Phi(p,Q)-\Phi(p,Q)(\gamma_5(Q-p)+\gamma_0m)
\hspace*{5cm}
$$
\be
+\gamma\int\frac{dk}{(p-k)^2}\left\{\Lambda_+(k)\Phi(p,Q)\Lambda_-(Q-k)-
\Lambda_+(p)\Phi(k,Q)\Lambda_-(Q-p)\right.
\label{matrix}
\ee
$$
\hspace*{5cm}\left.-\Lambda_-(k)\Phi(p,Q)\Lambda_+(Q-k)+\Lambda_-(p)\Phi(k,Q)\Lambda_+(Q-p)\right\},
$$
where we used projectors (\ref{projectors}) and the matrix wave function is
parametrized as
\be
\Phi(p,Q) = T(p)\left(\frac{1+\gamma_0}{2}\gamma_5\vph_+(p,Q)+
\frac{1-\gamma_0}{2}\gamma_5\vph_-(p,Q)\right)T^+(Q-p).
\label{phibig}
\ee

When written in components, equation (\ref{matrix}) readily reproduces the bound-state
equation (\ref{BG}) for the functions $\vph_{\pm}$ derived earlier in the 
framework of the Hamiltonian approach to the model.

In conclusion we give the connection between the meson-quark-antiquark ver\-tex 
$\Gamma(p,P)$ and the matrix wave function $\Phi(p,P)$:
\be
\Gamma(p,P)=\int\frac{dk}{2\pi}\gamma_0\frac{\Phi(k,P)}{(p-k)^2}\gamma_0,\quad
\bar \Gamma(p,P)=\gamma_0\Gamma^+(p,P)\gamma_0,
\label{Gamma}
\ee
where the vertices $\Gamma$ and $\bar{\Gamma}$ stand for the incoming and
the outgoing mesons, respectively. It is easy to check that, 
with such a definition, relation (\ref{Phidef}) is satisfied automatically (see also
Appendix A for the properties of $\Gamma$ and $\bar{\Gamma}$).

\subsection{Truncated Hilbert space and the problem of Hermiticity}

In this subsection we discuss properties of the matrix bound-state 
equation (\ref{matrix}), but let us make a comment concerning its scalar form
(\ref{BG}) first. As stated before, the norm of the
wave functions $\vph_{\pm}$ (\ref{norms}) is defined in an unusual way. Indeed,
the sign minus between its \lq\lq $+$" and \lq\lq $-$" parts appears quite naturally in the
Bogoliubov-like approach developed above, but it looks somewhat surprisingly in
the context of the standard Hamiltonian technique. Besides, it is easy to 
check that, if
the matrix bound-state equation (\ref{matrix}) is written in the Schr{\" o}dinger-like
form,
\be
Q^n_0\left(\vph_{+}^n\atop \vph_{-}^n \right)=
\hat {\cal H}\left(\vph_{+}^n\atop \vph_{-}^n \right),
\label{calH}
\ee
then the corresponding Hamiltonian $\hat {\cal H}$ appears non-Hermitian.
The following two questions should be discussed in this connection: i) what is
the reason for this, and ii) whether this does not lead to a disaster and the
eigenenergies of this equation are still real.

The answer to the first question becomes clear if one notices that the matrix
wave function (\ref{phibig}) satisfies the following conditions:
\be
\Lambda_+(p)\Phi(p,Q)\Lambda_+(Q-p)=\Lambda_-(p)\Phi(p,Q)\Lambda_-(Q-p) =0,
\label{ll}
\ee
so the phase space is truncated and the Hamiltonian 
$\hat {\cal H}$ acts in a subspace, that explains also the 
distorted norm (\ref{norms}).

In the meantime, the second question concerning the spectrum persists.
In order to answer it, let us integrate both sides of (\ref{matrix}) over $p$, do
the same for the complex conjugated equation, and take an appropriate linear
combination. Then one arrives at the relation
\be
\sum_{n=-\infty}^{+\infty}\left(Q^n_0-Q^{m*}_0\right)\int\frac{dp}{2\pi}
\left(\vph_+^n(p,Q)\vph_+^{m}(p,Q)-\vph_-^n(p,Q)\vph_-^{m}(p,Q)\right)=0,
\label{QQ}
\ee
which immediately leads to the following two conclusions:
\be
Q^n_0 = Q^{n*}_0,
\ee
and
\be
\begin{array}{rcl}
\ds\int\frac{\ds dp}{\ds 2\pi}\left(\vph_+^n(p,Q)\vph_+^{m}(p,Q)-\vph_-^n(p,Q)\vph_-^m(p,Q)
\right)&=&\delta_{nm},\\
&&\\
\ds\int\frac{dp}{2\pi}\left(\vph_+^n(p,Q)\vph_-^{m}(p,Q)-\vph_-^n(p,Q)\vph_+^m(p,Q)
\right)&=&0.
\label{norms_new}
\end{array}
\ee

It was already mentioned before (subsection \lq\lq Chiral condensate" above) 
that solutions of the system (\ref{BG}) appear in pairs: 
for each eigenvalue
$Q^n_0$ with the eigenfunction $(\vph^n_+,\vph^n_-)$ there exists another
eigenvalue, $-Q^n_0$, with the eigenfunction $(\vph^n_-,\vph^n_+)$. With
this symmetry, equation (\ref{norms_new}) can be rewritten in the form 
(\ref{norms}) 
where only positive eigenvalues enter\footnote{If not stated explicitly, we use the symbol
$\sum_n$ for summation over positive $n$'s only.}. Similarly in attempts to construct the
Green's function for the system (\ref{BG}) the completeness (\ref{complet}) 
can be derived.

Let us introduce operators $\hat{C}$ and $\hat{S}$: 
\be
\hat{C}(\hat{S})F(p,P)\equiv \gamma\int\frac{dk}{(p-k)^2}C(S)(p,k,P)F(k,P)
\ee
for an arbitrary function $F(p,P)$ with $C(p,k,P)$ and $S(p,k,P)$
defined in (\ref{cs}). 

Then the matrix Hamiltonian $\hat {\cal H}$ can be written in the form
\be
\hat {\cal H}=\left(
\begin{array}{cc}
K-\hat{C}&\hat{S}\\
-\hat{S}&-K+\hat{C}
\end{array}
\right)
=\gamma_0(K-\hat{C})+\gamma_1\hat{S},
\label{Hmat2}
\ee
where $K\equiv E(p)+E(P-p)$ is the kinetic energy. This is the term proportional to
$\gamma_1$ in (\ref{Hmat2}), which makes the Hamiltonian $\hat {\cal H}$ non-Hermitian.
The symmetry property of the solution
with respect to interchange of the plus and the minus components of the wave function discussed 
above follows immediately from the fact that $\hat {\cal H}$ anticommutes with $\gamma_5$,
so that if $\psi=\left(\vph_{+}\atop \vph_{-} \right)$ is the eigenfunction
corresponding to the eigenvalue $Q_0$, then $\psi'=\left(\vph_{-}\atop \vph_{+}
\right)=\gamma_5\psi$ is also a solution with the eigenvalue $-Q_0$,
\be
\hat {\cal H}\psi'=\hat {\cal H}\gamma_5\psi=-\gamma_5\hat {\cal H}\psi=-\gamma_5Q_0\psi=
-Q_0\psi'.
\ee 

Moreover, the eigenstate problem (\ref{calH}) for the operator $\hat {\cal H}$ can be 
formulated now in the form of an effective Dirac-type equation,
\be
\left[\gamma_0(K-\hat{C})+\gamma_1\hat{S}-Q_0\right]\psi=0,
\label{HDir}
\ee
where, as before,
\be
\psi=\left(\vph_{+}\atop \vph_{-} \right).
\ee

Mapping of the quark-antiquark bound-states problem to the properties of the
fermionic-type
equation (\ref{HDir}) may be continued, which is, however, beyond the scope of the present
paper.

\subsection{The chiral properties of the model in the matrix approach}
\label{cpm2}

In this subsection we return to the chiral properties of the 't~Hooft model and discuss
some of them in the framework of the matrix formalism.

In Appendix A we collect formulae useful for various calculations in the
suggested approach. They are entirely based on the definition of the matrix
wave function (\ref{phibig}) and properties of the matrix bound-state equation
(\ref{matrix}).

\subsubsection{The Gell-Mann-Oakes-Renner relation and the mass of the pion}\label{gmorr}

In order to demonstrate how the matrix approach works in practice let us derive the 
Gell-Mann-Oakes-Renner relation for the 't~Hooft model. We slightly relax the 
chiral limit introducing a small quark mass $m$. The matrix bound-state equation
(\ref{matrix})
is the main object of investigation now. We multiply it by
$\gamma_0\gamma_5$, take trace 
over spinor indices and integrate both sides of the resulting equation 
over the momentum $p$. A number of terms containing projectors $\Lambda_{\pm}$ 
disappears and the result reads
\be
Q_0\int\frac{dp}{2\pi}Sp[\gamma_0\gamma_5\Phi(p,Q)]-
Q\int\frac{dp}{2\pi}Sp[\gamma_1\gamma_5\Phi(p,Q)]=
-2m\int\frac{dp}{2\pi}Sp[\gamma_5\Phi(p,Q)].
\label{oak}
\ee

If one uses the definition of the matrix wave function and substitutes the pion 
solution (\ref{pion}) into it, then relation (\ref{oak}) simplifies even more
and takes the form:
\be
f_{\pi}^2M_{\pi}^2=-2m\langle\bar{q}q\rangle,
\label{glm}
\ee
in which one can easily recognize the celebrated Gell-Mann-Oakes-Renner relation
\cite{GOR}. This defines the mass of the pion near the chiral limit,
\be
M_{\pi}^2=2m\int_0^{\infty}dp\cos\theta(p).
\ee

With the help of the numerical solution for $\theta$ (see Fig.\ref{figtheta})
and the footnote at page \pageref{exactcon} one can find:
\be
M_{\pi}^2=\sqrt{\frac{2\pi^2m^2\gamma}{3}}\sim m\sqrt{\gamma}.
\label{Mpi}
\ee

Note that in the case of the chirally invariant vacuum, {\it i.e.}, for the 
analytic solution (\ref{badsol}), equation (\ref{glm}) is trivial as its
both sides vanish simultaneously.

\subsubsection{The pionic solution beyond the chiral limit}

\begin{figure}[t]
\centerline{\hspace*{0.5cm}\epsfig{file=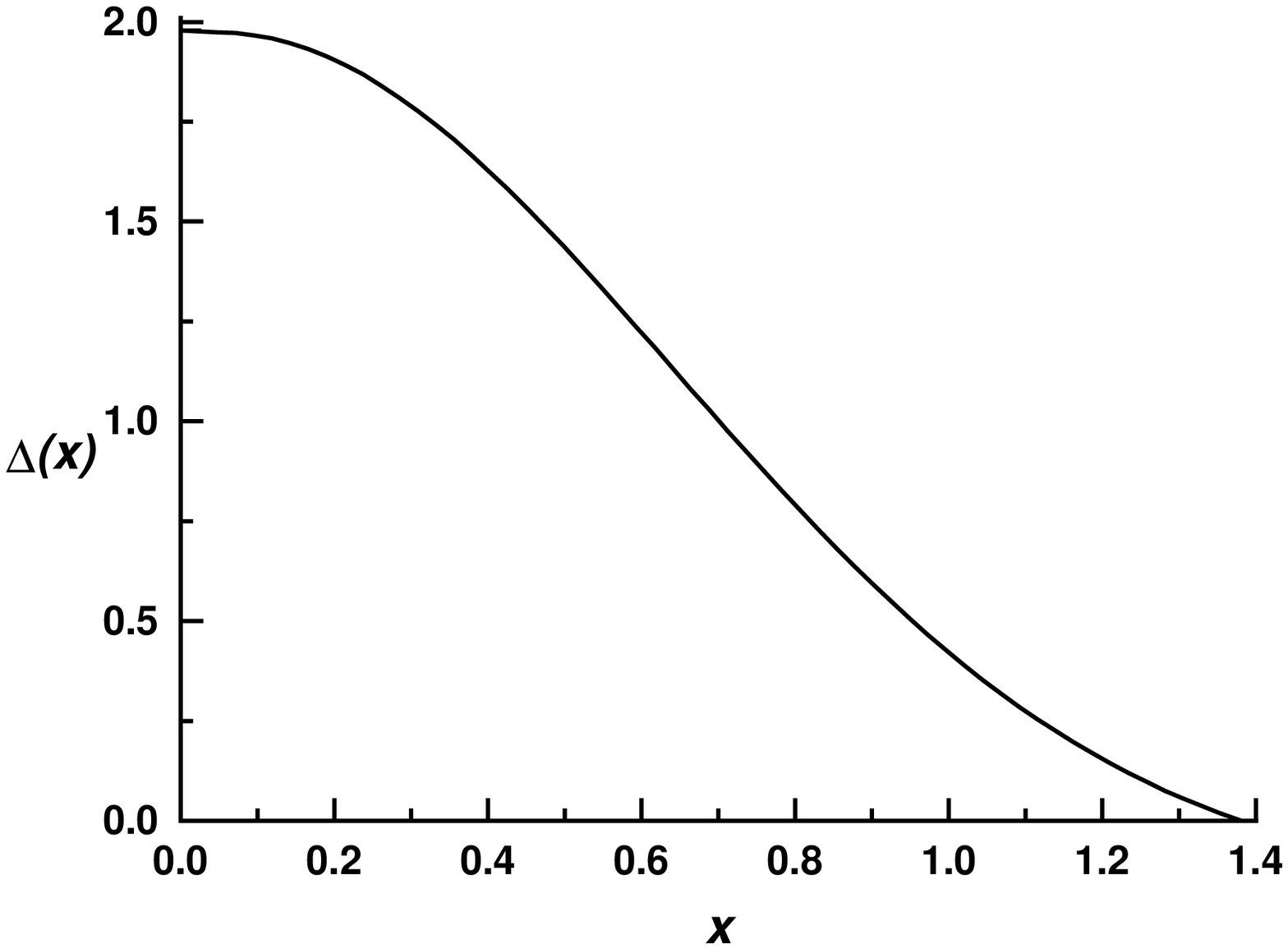,width=9cm}\hspace{-1cm}
            \epsfig{file=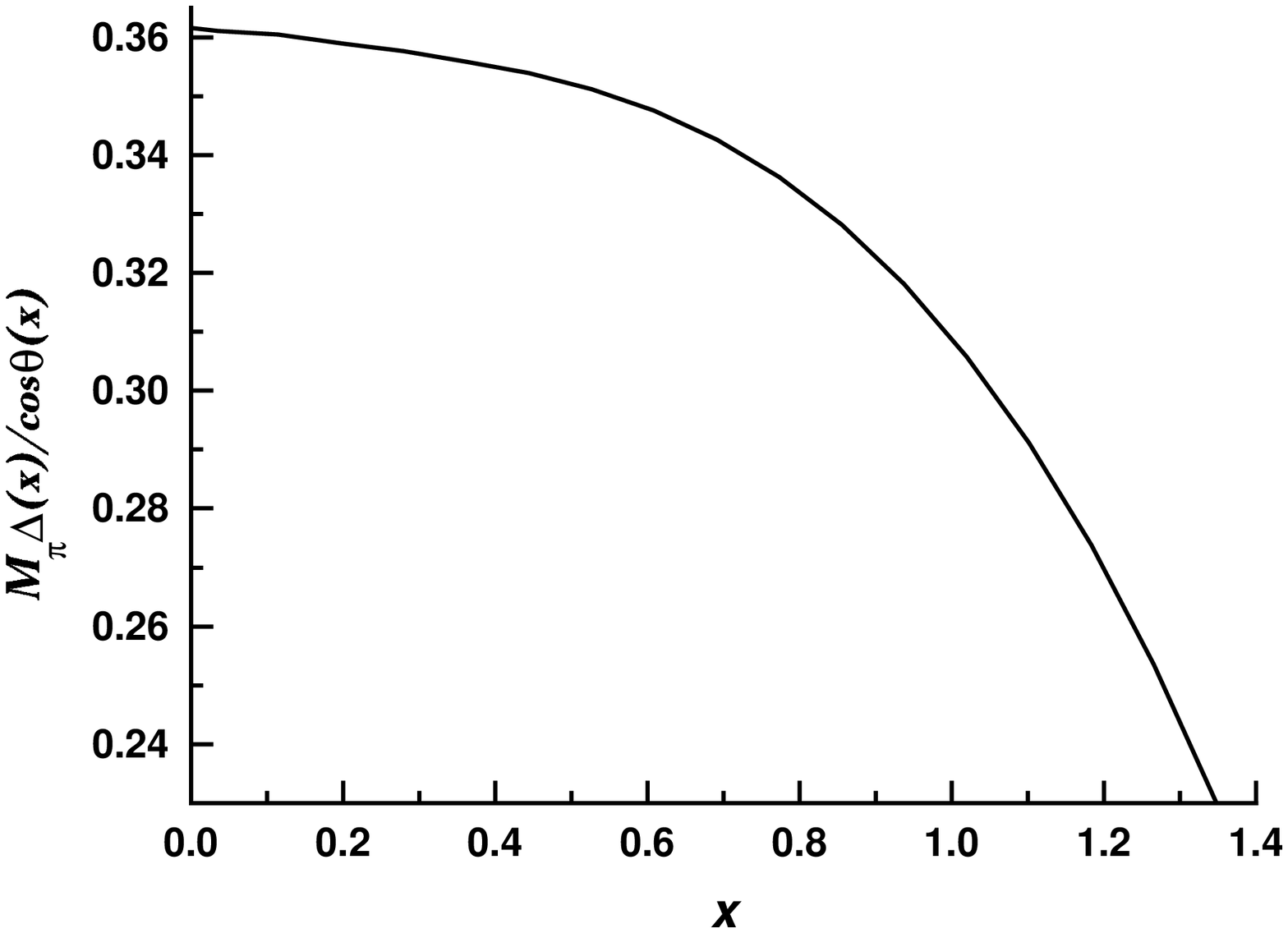,width=9cm}}
\caption{Numerical solution for the function $\Delta$ and the ratio of the subleading term
in (\ref{pisol2}) to the leading one for $m=0.18$. As before, 
variable $x$ comes from the change $p=\tan(x)$, all dimensional quantities are given in proper 
units of $(2\gamma)^{1/2}$.}\label{figdelta}
\end{figure}

With the pion mass (\ref{Mpi}) in hands we are in the position to go slightly
beyond the chiral limit and to study the pionic solution in the rest frame. A simple
analysis demonstrates that in the pion rest frame the functions $g_{\pi}(p,P)$ and 
$f_{\pi}(p,P)$ defined in (\ref{gfdef}) behave like
\be
g_{\pi}(p,P=0)\sim\frac{1}{\sqrt{M_{\pi}}}\cos\theta(p)+O\left(M_{\pi}^{3/2}\right),\quad
f_{\pi}(p,P=0)\sim O\left(\sqrt{M_{\pi}}\right).
\ee

Therefore, the leading correction to the pion wave function comes from $f_{\pi}(p,P)$
and $\varphi_{\pm}^{\pi}(p,P=0)$ can be parametrized in the form (see (\ref{pion})):
\be
\varphi_{\pm}^{\pi}(p,P=0)={\tilde N}_{\pi}^{-1}\left[\frac{\cos\theta(p)}
{\sqrt{M_{\pi}}}\pm\sqrt{M_{\pi}}\Delta(p)\right],
\label{pisol2}
\ee
where we have extracted the explicit dependence of the coefficients on the small pion
mass, so that the unknown correction function $\Delta(p)$ (note that $\Delta(p)$ has the same parity
as $\cos\theta(p)$, $i.e.$, it is even) does not depend on $M_{\pi}$.
The dimensionless norm ${\tilde N}_{\pi}$ is
\be
{\tilde N}_{\pi}^2=4\int\frac{dp}{2\pi}\Delta(p)\cos\theta(p),
\ee
providing the correct normalization for $\varphi_{\pm}^{\pi}$.

Substituting (\ref{pisol2}) into (\ref{BG}), one arrives at a system of two equations, the
first of which is satisfied identically due to (\ref{system}), whereas the other one defines 
the correction function $\Delta(p)$,
\be
\frac12\cos\theta(p)-p\Delta(p)\sin\theta(p)=\frac{\gamma}{2}\vpint\frac{dk}{(p-k)^2}[\Delta(p)-
\Delta(k)]\cos[\theta(p)-\theta(k)].
\label{eqD}
\ee

Equation (\ref{eqD}) is subject to numerical studies which are beyond the scope of the
present paper. 

Exact numerical solutions for the pion wave functions in the rest frame for several values
of the quark mass taken from \cite{Ming Li} are given in Fig.\ref{figphi}, so that the
function $\Delta(p)$ can be extracted directly from these data. In Fig.\ref{figdelta} we
give the form of the function $\Delta(p)$ and the ratio of the
correction defined by $\Delta(p)$ to the leading term in (\ref{pisol2}) 
for $m=0.18$ (in units of $\sqrt{2\gamma}$). 
One can see from the right plot in Fig.\ref{figdelta} that for the given
value of the
quark mass the correction is of order one third at largest and decreases with the growth of the
argument. 

\subsubsection{The pion decay constant revisited}

\begin{figure}[t]
\begin{center}
\epsfxsize=10cm
\epsfbox{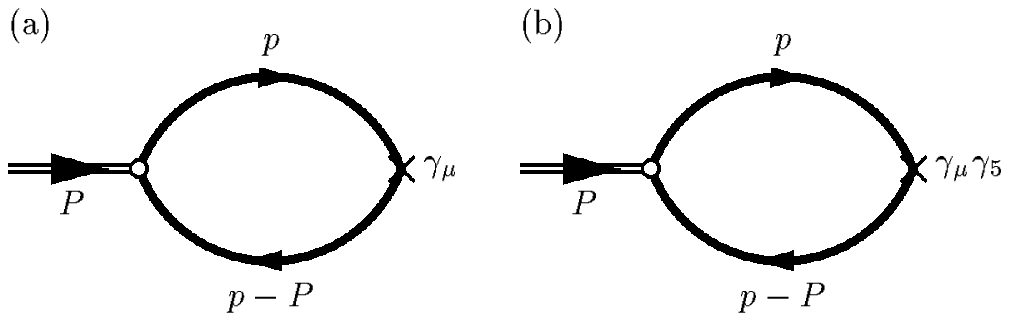}
\caption{The meson-vector (figure (a)) and meson-axial-vector (figure (b)) 
current couplings.}\label{figcur}
\end{center}
\end{figure}

In this subsection we give another example of calculations using the matrix approach ---
namely, we calculate the pion decay constant once again, which comes now from the
fish-like diagram depicted in Fig.\ref{figcur}(b).

The matrix element written for this diagram reads
\be
A_{\mu}=\frac{-i\gamma}{\sqrt{N_C}}N_C(-1)\int\frac{d^2p}{(2\pi)^2}Sp
\left[S(p)\Gamma_n(p,P)S(p-P)\gamma_{\mu}\gamma_5\right],
\label{Acur}
\ee
where the factor $-i\gamma/\sqrt{N_C}$ comes from the vertex, whereas $N_C$ and
(-1) are due to the fermionic loop. Working out the
integral over $p_0$ and using relations (\ref{rel1}) and (\ref{rel2}), one arrives
at a simple formula,
\be
A_{\mu}=\sqrt{N_C}\int\frac{dp}{2\pi}Sp[\Phi_n(p,P)\gamma_{\mu}\gamma_5].
\ee

Then, on substituting the explicit form of the matrix wave function $\Phi_n$, putting 
$m=0$, and, finally,
on taking the integral over $p$ by means of the orthogonality condition (\ref{norms}),
one finds this matrix element in the chiral limit to be
\be
A_{\mu}=\delta_{n\pi}\sqrt{\frac{N_C}{\pi}}P_{\mu}\frac{1}{\sqrt{2P_0}}.
\label{Mmu}
\ee

Comparing expression (\ref{Mmu}) with the definition of the decay constant for
the $n$th meson,
\be
A_{\mu}=f_nP_{\mu}\Psi_n(0),
\ee
one can easily conclude that in the chiral limit
\be
f_n=\delta_{n\pi}\sqrt{\frac{N_C}{\pi}},
\ee
that coincides with relation (\ref{fpi}).

\subsection{Quark-quark scattering amplitude}
\label{GGG}
\begin{figure}[t]
\begin{center}
\epsfxsize=14cm
\epsfbox{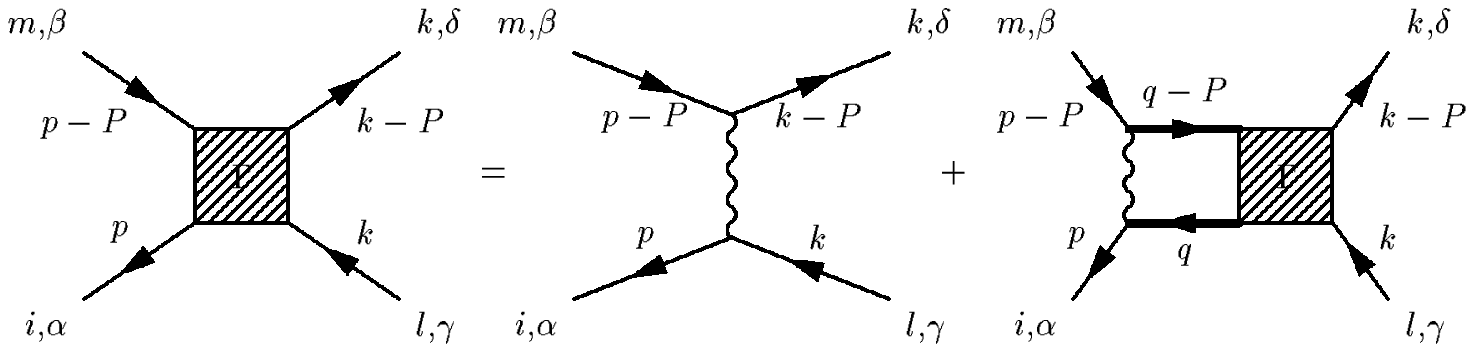}
\caption{Graphical representation of the quark-quark scattering amplitude.}\label{figGamma}  
\end{center}
\end{figure}

In this subsection we come to calculation of one of the most fundamental objects
in the theory --- the quark-quark scattering amplitude. If known, this object allows one
to define the dressed current vertices and thus to investigate such properties
of the theory as Ward identities, current conservation laws,
{\it etc}. Equation for this object is given in the diagrammatic form in
Fig.\ref{figGamma} and reads
\be
\Gamma_{ik,lm}^{\alpha\delta,\gamma\beta}(p,k,P_{\mu})=
-ig^2\left(\gamma_0\right)_{il}\left(\gamma_0\right)_{km}\frac{1}{(p-k)^2}
\left(t^a\right)_{\gamma}^{\alpha}\left(t^a\right)_{\beta}^{\delta}
\ee
$$
-ig^2\int\frac{d^2q}{(2\pi)^2}\frac{1}{(p-q)^2}
\left(\gamma_0\right)_{is}S_{st}(q)\Gamma_{tk,ln}^{\omega\delta,\gamma\sigma}
(q,k,P_{\mu})S_{nr}(q-P)\left(\gamma_0\right)_{rm}
\left(t^a\right)_{\omega}^{\alpha}\left(t^a\right)_{\beta}^{\sigma},
$$
where Greek and Latin letters stand for the colour and the spinor indices, respectively.

Then one can separate the colour structure of $\Gamma$,
\be
\Gamma_{ik,lm}^{\alpha\delta,\gamma\beta}=\frac{1}{N_C}\delta_{\beta}^{\alpha}
\delta_{\gamma}^{\delta}\Gamma_{ik,lm},
\ee
and it is also convenient to introduce a new function $\tilde{\Gamma}$ defined as
\be
\tilde{\Gamma}_{sk,lr}(q_{\mu},k,P_{\mu})=S_{st}(q)\Gamma_{tk,ln}(q,k,P_{\mu})
S_{nr}(q-P).
\ee

Due to the instantaneous type of the interaction induced by the two-dimensional gluon, it is
also useful to define the amplitude $\Phi$ as an integral of $\tilde{\Gamma}$ 
over $q_0$,
\be
\int\frac{dq_0}{2\pi}\tilde{\Gamma}_{sk,lr}(q_{\mu},k,P_{\mu})=
\Phi_{sk,lr}(q,k,P_{\mu}).
\ee

Then the equation for the new function is
\be
\Phi_{ak,lb}(p,k,P_{\mu})=\frac{(2\pi)^2\gamma}{(p-k)^2}(\gamma_0)_{il}
(\gamma_0)_{km}\left[i\int\frac{dp_0}{2\pi}S_{ai}(p)S_{mb}(p-P)\right]
\label{eqG}
\ee
$$
+\gamma\int\frac{dq}{(p-q)^2}(\gamma_0)_{is}(\gamma_0)_{rm}
\Phi_{sk,lr}(q,k,P_{\mu})\left[i\int\frac{dp_0}{2\pi}S_{ai}(p)S_{mb}(p-P)\right].
$$

The object in the square brackets can be easily integrated out using formula
(\ref{SS}) and the result reads
$$
i\int\frac{dp_0}{2\pi}S_{ai}(p)S_{mb}(p-P)=
\frac{(\Lambda_+(p)\gamma_0)_{ai}(\Lambda_-(p-P)\gamma_0)_{mb}}{E(p)+E(P-p)-P_0}
\hspace*{2cm}
$$
\be
\hspace*{4cm}+\frac{(\Lambda_-(p)\gamma_0)_{ai}(\Lambda_+(p-P)\gamma_0)_{mb}}{E(p)+E(P-p)+P_0}.
\ee

The simplest way to proceed further is to guess the general structure of the
amplitude $\Phi$ to be
\be
\Phi_{sk,lr}(q,k,P_{\mu})=\sum_{n=-\infty}^{\infty}\frac{\Phi_{sr}^n(q,P)\chi_{kl}^n(k,P)}{P_0-P_0^n},
\ee
so that after some algebraic transformations with the use of the matrix
bound-state equation (\ref{matrix}) one arrives at
\be
\sum_{n=-\infty}^{\infty}\chi_{kl}^n(k,P)\left[
\frac{\Lambda_+(p)\Phi^n(p,P)\Lambda_-(P-p)}{E(p)+E(P-p)-P_0}
-\frac{\Lambda_-(p)\Phi^n(p,P)\Lambda_+(P-p)}{E(p)+E(P-p)+P_0}
\right]_{ab}
\label{eqchi}
\ee
$$
=\frac{(2\pi)^2\gamma}{(p-k)^2}\left[
\frac{(\Lambda_+(p))_{al}(\Lambda_-(P-p))_{kb}}{E(p)+E(P-p)-P_0}
-\frac{(\Lambda_-(p))_{al}(\Lambda_+(P-p))_{kb}}{E(p)+E(P-p)+P_0}
\right].
$$

We parametrize the function $\chi^n(p,P)$ in the form:
\be
\chi^n(p,P)=2\pi\gamma\int dq\frac{\psi^n(q,P)}{(p-q)^2},
\label{psidef}
\ee
with $\psi^n(p,P)$ being a new unknown function. Substituting (\ref{psidef}) into
(\ref{eqchi}), we arrive at two simple equations defining $\psi^n(p,P)$,
\be
\sum_{n=-\infty}^{\infty}Sp\left[\psi^n(q,P)\Phi^n_{\pm}(p,P)\right]=\pm 2\pi\delta(p-q).
\ee
With the help of the formulae from Appendix A the following solution to these equations
can be found:
\be
\psi^n(p,P)={\rm sign}(n)\Phi^{n+}(p,P),
\ee 
and, hence,
\be
\chi^n(p,P)=2\pi\gamma {\rm sign}(n)\int dq\frac{\Phi^{n+}(q,P)}{(p-q)^2},
\ee
that gives for the quark-quark scattering amplitude \cite{decay}:
$$
\Gamma_{ik,lm}(p,k,P_{\mu}) = \frac{2\pi i\gamma}{(p-k)^2}(\gamma_0)_{il}
(\gamma_0)_{km}
-i\gamma^2\sum_n\frac{1}{P_0-P^n_0}(\Gamma_n(p,P))_{im}(\bar\Gamma_n(k,P))_{kl}
$$
\be
+i\gamma^2\sum_n\frac{1}{P_0+P^n_0}(\gamma_0\bar\Gamma_n(P-p,P)\gamma_0)_{im}  
(\gamma_0\Gamma_n(P-k,P)\gamma_0)_{kl},
\label{solGamma}
\ee
where the sum over mesons $\sum_n$ counters only positive excitation numbers
with $n>0$.

A couple of comments concerning the solution (\ref{solGamma}) is in order.
First of all, note that, once the wave function $\Phi^n$ contains only symmetric
matrices $\gamma_0$ and $\gamma_5$, then $\Phi^n_{ab}=\Phi^n_{ba}$ and, hence,
\be
\Gamma_{ik,lm}(p,k,P_{\mu})=\Gamma_{ml,ki}(p,k,P_{\mu}).
\ee

The other comment concerns the form of the solution (\ref{solGamma}). One can see
that the
quark-quark scattering goes through the formation of a one-meson intermediate state and
that the last two terms in (\ref{solGamma}) give nothing but the sum over the full 
mesonic propagators.

Such way, we end with an effective diagrammatic technique which involves the dressed quark
propagator (\ref{S}) (see (\ref{SS}) for its ultimate form), the dressed meson-quark-antiquark
amplitude ($\Gamma$ for the incoming and $\bar{\Gamma}$ for the outgoing mesons
given by (\ref{Gamma})), the dressed quark-antiquark scattering amplitude (\ref{solGamma}) and
the constant $-i\gamma/\sqrt{N_C}$ which is to be inserted into every vertex where a meson
couples to a quark-antiquark pair. In addition, each quark loop brings two extra factors,
the standard fermionic (-1) and $N_C$ from the colour trace. One can use these ingredients
as bricks for building any hadronic process in the theory.

\begin{figure}[t]
\begin{center}
\epsfxsize=6cm
\epsfbox{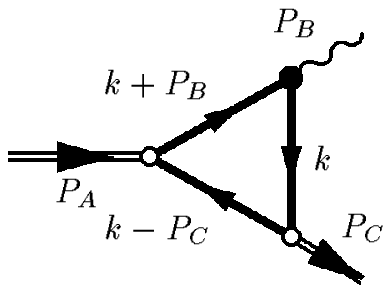}
\caption{Graphical representation of the mesonic form factor.}\label{figff}
\end{center}
\end{figure}

\subsection{Vector and axial-vector currents conservation. Ward identities.}

As stated in the previous subsection, with the quark-quark scattering amplitude
in hands we are in the position to study properties of currents in the
't~Hooft model. Let us prove the currents conservation first. For the vector
current we are interested in the matrix element
\be
V^M_{\mu}(P)=\langle \Omega|\bar q\gamma_{\mu} q|M,P\rangle,
\ee
depicted in Fig.\ref{figcur}(a).

On writing the corresponding matrix element and performing the integration over
the energy $p_0$, one finds:
\be
V^M_{\mu}=i\gamma\sqrt{N_C}\int\frac{d^2p}{(2\pi)^2}Sp[S(p-P)\gamma_{\mu}S(p)\Gamma_M(p,P)]= 
\sqrt{N_C}\int \frac{dp}{2\pi} Sp[\gamma_{\mu}\Phi_M(p,P)].
\label{V1}
\ee

It is easy to check that, multiplying the bound-state equation (\ref{matrix}) 
by $\gamma_0\sqrt{N_C}/2\pi$, taking trace over the spinor indices, and
integrating all terms of the resulting equation over the momentum $p$, one
arrives at the
relation:
\be
P_0^M\left[\sqrt{N_C}\int\frac{dp}{2\pi}Sp[\gamma_0\Phi_M(p,P)]\right]-
P\left[\sqrt{N_C}\int\frac{dp}{2\pi}Sp[\gamma_1\Phi_M(p,P)]\right]=0.
\ee
Combining it with the definition (\ref{V1}), one finds that the 
vector current is conserved,
\be
P_0^MV^M_0-PV^M=0.
\ee

Similarly, defining the axial-vector current as
\be
A^M_{\mu}(P)=\langle \Omega|\bar q\gamma_{\mu}\gamma_5 q|M,P\rangle,
\ee 
{\it i.e.}, using the diagram (b) in Fig.\ref{figcur} and performing the same steps
concerning the bound-state equation as before,
but with the evident change $\gamma_0\to\gamma_1$ in the multiplier,
one can derive the axial-vector current divergency in the form:
\be
P_0^MA^M_0-PA^M =-2m\sqrt{N_C}\int\frac{dk}{2\pi}Sp[\gamma_5\Phi_M],
\ee
which turns into the axial-vector-current conservation law in the chiral limit.

It is instructive to see how the same relations appear in the Hamiltonian
approach. We shall concentrate only on the vector current conservation law as
a similar analysis for the axial-vector current is straightforward then.

We start from the definition of the vector current,
\be
J_{\mu}(x)=\bar{q}(x)\gamma_{\mu}q(x),
\ee
and reformulate it in terms of mesonic creation and annihilation 
operators,\footnote{Note that we have restored the explicit dependence of operators $m_n$
on time in the form $m_n(x_0,P)=e^{-iP_0^nx_0}m_n(P)$ with $P_0^n$ being the energy of the
$n$th mesonic state moving with the total momentum $P$.}
\be
J_0(x)=2\sqrt{N_C}\int\frac{dP}{2\pi}e^{iPx}\sum_ne^{-iP_0^nx_0}m_n(P)
\int\frac{dp}{2\pi}g_n(p,P)f_0(p,P)+h.c.,
\ee
\be
J(x)=2\sqrt{N_C}\int\frac{dP}{2\pi}e^{iPx}\sum_ne^{-iP_0^nx_0}m_n(P)
\int\frac{dp}{2\pi}f_n(p,P)g_0(p,P)+h.c.,
\ee
where $f_n(p,P)$ and $g_n(p,P)$ are introduced in (\ref{gfdef}),
whereas $f_0$ and $g_0$ are other notations for the pion wave functions $f_{\pi}$ and $g_{\pi}$.

Then calculating the corresponding matrix element explicitly one finds:
\be
P_0^MV_0^M-PV^M=i\langle\Omega|\partial_{\mu}J_{\mu}(0)|M,P\rangle=\hspace*{4cm}
\ee
$$
=2\sqrt{N_C}\int\frac{dp}{2\pi}\left[P_0^Mg_M(p,P)f_0(p,P)-Pf_M(p,P)g_0(p,P)\right]=0,
$$
where the r.h.s. of this equation vanishes due to the bound-state equation
(\ref{BG}) or (\ref{matrix}).

Now we turn to the investigation of the mesonic form-factors defined by the
diagram depicted in Fig.\ref{figff}, but we need to know the dressed
current-quark-antiquark vertices first.

\begin{figure}[t]
\begin{center}
\epsfxsize=14cm
\epsfbox{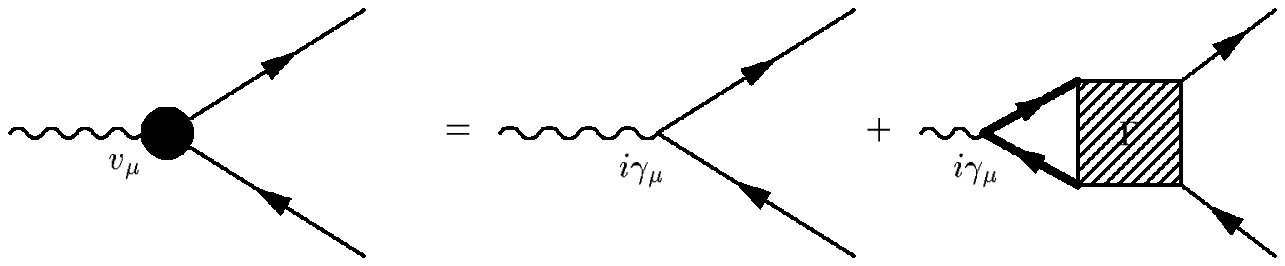}
\caption{Graphical representation of the dressed vector current-quark-antiquark
vertex.}\label{figvec}
\end{center}
\end{figure}

Let us start from the vector current. The corresponding expression for the
diagrams given in Fig.\ref{figvec} reads 
\be
\left(v_{\mu}^{kl}(p,P)\right)_{\beta}^{\alpha}=
i(\gamma_{\mu})_{kl}\delta_{\beta}^{\alpha}-
\int\frac{d^2q}{(2\pi)^2}S_{da}(q)i(\gamma_{\mu})_{ab}\delta_{\delta}^{\gamma}
S_{bc}(q-P)\Gamma_{ck,ld}^{\delta\alpha,\beta\gamma}(q,p,P)=
v_{\mu}^{kl}\delta_{\beta}^{\alpha},
\label{vv}
\ee
where, as before, the colour indices are denoted by the Greek symbols, whereas Latin ones stand
for the spinor indices. Substituting the explicit form of the quark-quark scattering 
amplitude $\Gamma_{ck,ld}^{\delta\alpha,\beta\gamma}(q,p,P)$ from (\ref{solGamma}) into 
(\ref{vv}), one finds \cite{decay}:
$$
v_{\mu}(p,P_{\mu})=i\gamma_{\mu}+i\gamma\sum_n \frac{\bar \Gamma_n(p,P)}{P_0-P^n_0}\int
\frac{dq}{2\pi} Sp[\gamma_{\mu}\Phi_n(q,P)]
$$
\be
-i\gamma\sum_n \frac{\gamma_0\Gamma_n(P-p,P)\gamma_0}{P_0+P^n_0}\int \frac{dq}{2\pi}
Sp[\gamma_{\mu}\Phi_n^+(P-q,P)].
\label{v}
\ee

After tedious but straightforward calculations using the explicit form of the
matrix bound-state equation (\ref{matrix}) and the matrix wave function 
(\ref{phibig}) the following relation can be proved:
\be
-iP_{\mu}v_{\mu}(p,P)=S^{-1}(p)-S^{-1}(p-P),
\label{divv}
\ee
which is nothing but the vector Ward identity for the 't~Hooft model
(similar expression in the light-cone gauge was derived in 
\cite{Einhorn}).

Now it is a simple task to substitute the solution (\ref{divv}) into the matrix
element written for the diagram in Fig.\ref{figff} with the vector current
instead of the curly line and to arrive at the vector current conservation law,
\be
Q_{\mu}\langle M,P|v_{\mu}|M',P'\rangle=0,\quad Q_{\mu}=P_{\mu}-P'_{\mu}.
\label{averv}
\ee

Similar calculations give the following results for the axial-vector current in
the chiral limit \cite{decay},
$$
a_{\mu}(p,P_{\mu})=i\gamma_{\mu}\gamma_5+i\gamma\sum_n \frac{\bar \Gamma_n(p,P)}{P_0-P^n_0}\int
\frac{dq}{2\pi} Sp[\gamma_{\mu}\gamma_5\Phi_n(q,P)]
$$
\be
-i\gamma\sum_n \frac{\gamma_0\Gamma_n(P-p,P)\gamma_0}{P_0+P^n_0}\int \frac{dq}{2\pi}
Sp[\gamma_{\mu}\gamma_5\Phi_n^+(P-q,P)],
\label{a}
\ee
\be
-iP_{\mu}a_{\mu}(p,P) = S^{-1}(p)\gamma_5+\gamma_5S^{-1}(p-P),
\label{diva}
\ee
and, finally,
\be
Q_{\mu}\langle M,P|a_{\mu}|M',P'\rangle=0,\quad Q_{\mu}=P_{\mu}-P'_{\mu}.
\label{avera}
\ee
Relation (\ref{diva}) plays the role of the axial-vector Ward identity.

In conclusion let us notice that one could arrive at the same results using the
Hamiltonian approach but at the price of a much more complicated algebra and a much
less transparent interpretation of the results in terms of Feynman diagrams. 

\subsection{The pionic vertex}

The general structure of the mesonic vertex can be considerably simplified in
case of the pion, since the explicit form of the pionic wave function is known. Indeed,
substituting the solution (\ref{pion}) into the matrix form (\ref{phibig})
and then into the definition (\ref{Gamma}), one easily finds \cite{decay}:
\be
\Gamma_{\pi}(p,P)=S^{-1}(p)(1+\gamma_5)-(1-\gamma_5)S^{-1}(p-P).
\label{pivert1}
\ee

It is also instructive to compare formula (\ref{pivert1}) with the Ward
identities for the vector and axial-vector currents derived in the previous
subsection, equations (\ref{divv}) and (\ref{diva}), respectively.
As a result, one finds the following relation for the pionic vertex \cite{decay}:
\be
\Gamma_{\pi}(p,P)=-iP_{\mu}v_{\mu}(p,P)-iP_{\mu}a_{\mu}(p,P).
\label{pivert}
\ee

Note that it is not surprise that the pion couples not only to the axial-vector,
but to the vector current as well. The reason is that in the two-dimensional
theory the axial-vector and the vector currents are dual to one another,
\be
J_{\mu}^5(x)=\varepsilon_{\mu\nu}J^{\nu}(x),
\ee
where $\varepsilon_{\mu\nu}$ is the totally-antisymmetric Levy-Civita tensor
in two dimensions.

\section{Strong decays}

This section is devoted to investigation of hadronic processes in the 't~Hooft model at
the example of the decay $A\to B+C$. In subsections 4.1 and 4.2 we derive the amplitude of
such a decay using the Hamiltonian and matrix approach, respectively. We discuss its
properties and correspondence with nonrelativistic models. Subsection 4.3 is devoted
to derivation and justification of the two-dimensional Adler selfconsistency condition
(\lq\lq Adler zero").

\subsection{Suppressed terms in the Hamiltonian}

In section 2 we developed the Hamiltonian approach to QCD$_2$ and
diagonalized the Hamiltonian of the model in the mesonic sector. Now let us turn
to corrections to the Hamiltonian (\ref{HHHH}) suppressed by powers of
$N_C$ in the denominator. The leading correction is of order $O(1/\sqrt{N_C})$ 
and it comes from the terms containing the products $MB$, $MD$, $M^+B$, and 
$M^+D$ of the operators introduced in (\ref{operators}),
$$
\Delta H=-\frac{\gamma}{\sqrt{N_C}}\int\frac{dpdkdqdQ}{(2\pi)^3(p-k)^2}
\cos\frac{\theta(p)-\theta(k)}{2}\sin\frac{\theta(Q-p)-\theta(Q-k)}{2}
$$
\be
\times\left[M^+(p,p-Q)M^+(k-Q,q)M(k,q)+M^+(p-Q,p)M^+(q,k-Q)M(q,k)\right.
\label{dH}
\ee
$$
\left.-M^+(q,p)M(q,p-Q)M(k-Q,k)-M^+(p,q)M(p-Q,q)M(k,k-Q)\right],
$$
where $M$ and $M^+$ can be defined through the meson creation and
annihilation operators (see also (\ref{mM})),
\be
\begin{array}{ccc}
M^+(p,k)&=&\sum\limits_{n=0}^{\infty}\left[m_n^+(k-p)\vph_+^n(k,k-p)-m_n(p-k)\vph_+^n(p,p-k)\right],\\
{}\\
M(p,k)&=&\sum\limits_{n=0}^{\infty}\left[m_n(k-p)\vph_+^n(k,k-p)-m_n^+(p-k)\vph_+^n(p,p-k)\right].
\end{array}
\ee

As easily seen from (\ref{dH}), this correction describes vertices with three
mesons involved. 

\subsection{$A\to B+C$ decay amplitude}

Now we are in the position to study the hadronic processes in the 't~Hooft model.
Strong decays $A \ra B+C$ are of most interest for us \cite{decay}. 
As mentioned before, in the
leading order in $N_C$ the 't~Hooft model describes free mesons (see (\ref{HHHH})),
whereas the interaction between them is hidden in the suppressed terms partially 
restored in
the previous subsection. Thus we expect the amplitude $M(A \ra B+C)$ to be of order
$1/\sqrt{N_C}$, whereas $M(A+B\to C+D)\sim 1/N_C$. In this subsection we 
study the influence of the backward motion
described by the $\vph_-$ component of the mesonic wave function on the above mentioned
amplitude.

The Hamiltonian approach developed before gives the most straightforward way to
calculate the amplitude of the strong decay since one just needs to evaluate the
following matrix element:
\be
M(A \ra B+C)=\langle B(P_B)C(P_C)|H+\Delta H|A(P_A)\rangle =
\langle B(P_B)C(P_C)|\Delta H|A(P_A)\rangle.
\ee

With the help of the explicit form of the operator $\Delta H$ given in (\ref{dH}) one
easily finds the general form of the amplitude in terms of mesonic wave functions in
the rest frame of the decaying particle $A$ ($P_A=0$, $P_B=-P_C=p$) to be (see also
\cite{Swanson} where a six-term decay amplitude is discussed):
\be
M(A \ra B+C)=\hspace*{7cm}
\label{ampl}
\ee
$$
\frac{\gamma}{\sqrt{N_C}}\int\frac{dkdq}{(q-k)^2}\left\{\right.
-\varphi_-^A(k+p,0)\varphi_-^B(k+p,0)[c(-p,q,k)\varphi_+^C(q,-p)+s(-p,q,k)\varphi_{-C}]
$$
$$
-\varphi_+^A(k+p,0)\varphi_+^C(k,-p)[c(p,q+p,k+p)\varphi_+^B(q+p,p)+s(p,q+p,k+p)\varphi_-^B(q+p,p)]
$$
$$
-\varphi_+^C(k,-p)\varphi_-^B(k+p,p)[s(0,q+p,k+p)\varphi_+^A(q+p,0)+c(0,q+p,k+p)\varphi_-^A(q+p,0)]
$$
$$
+\varphi_-^C(k,-p)\varphi_+^B(k+p,p)[c(0,q+p,k+p)\varphi_+^A(q+p,0)+s(0,q+p,k+p)\varphi_-^A(q+p,0)]
$$
$$
+\varphi_-^A(k+p,0)\varphi_-^C(k,-p)[s(p,q+p,k+p)\varphi_+^B(q+p,p)+c(p,q+p,k+p)\varphi_-^B(q+p,p)]
$$
$$
+\varphi_+^A(k+p,0)\varphi_+^B(k+p,p)[s(-p,q,k)\varphi_+^C(q,-p)+c(-p,q,k)\varphi_-^C(q,-p)]\left.\right\}
$$
$$
\hspace*{7cm}+(B \leftrightarrow C, p \leftrightarrow -p),
$$
where 
$$
c(p,q,k)=\cos\frac{\theta(k)-\theta(q)}{2}\sin\frac{\theta(p-k)-\theta(p-q)}{2},
$$
$$
s(p,q,k)=\sin\frac{\theta(k)-\theta(q)}{2}\cos\frac{\theta(p-k)-\theta(p-q)}{2}.
$$

In spite of its frightening appearance, the amplitude (\ref{ampl}) has a very simple
structure. Indeed, it contains six terms, {\it i.e.}, three times more than one could
naively expected and this is entirely due to the presence of the $\vph_-$ component
in the mesonic wave function. 
If one neglects the backward motion contributions in the amplitude (\ref{ampl})
and inserts the nonrelativistic values of the angle $\theta$
($\cos\theta(p)=1,\;\sin\theta(p)=p/m$), then it reproduces the
standard quark-model decay amplitude due to the OGE Coulomb interaction
\cite{Barnes} adapted to the two-dimensional case. It is clear, however, that the substitution
of the nonrelativistic angle is not justified for kinematical reasons. 

\begin{figure}[t]
\begin{center}
\epsfxsize=14cm
\epsfbox{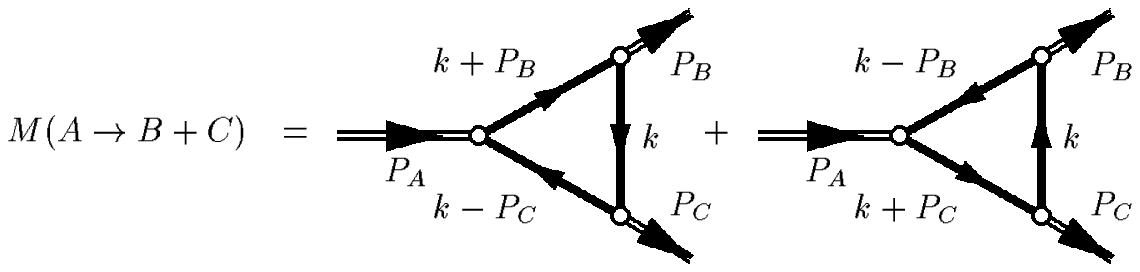}
\caption{Graphical representation of the amplitude for the decay $A\to B+C$.}\label{figampl}
\end{center}
\end{figure}

One can arrive at exactly the same expression for the amplitude using the matrix
approach and writing the matrix element for the diagrams depicted in
Fig.\ref{figampl},
$$
M(A \ra B+C)=\hspace*{7cm}
$$
\be
-\frac{i\gamma^3}{\sqrt{N_C}}\int \frac{d^2k}{(2\pi)^2} Sp[\Gamma_A(k+P_B,P_A)S(k-P_C)\bar \Gamma_C(k,P_C)
S(k)\bar \Gamma_B(k+P_B,P_B)S(k+P_B)]
\label{ABC}
\ee
$$
\hspace*{7cm}+(B \leftrightarrow C).
$$

In the meantime, the matrix approach proves more convenient in studies of the decay
amplitude (\ref{ampl}), (\ref{ABC}), so we stick with it in the next subsection 
considering pions in the final state.

\subsection{Adler selfconsistency condition}

In this subsection we have one more careful look at the pions --- namely, at their role in the
hadronic processes. We remind the reader that these are pions to suffer most drastically from the
presence of the backward motion of the $q\bar q$ pair described by the $\varphi_-$
component of the wave function. On the other hand, hadronic processes with the pions in the
final state are much better investigated experimentally, so that any theoretical hint as
to
how the chiral nature of the pion affects hadronic decays is of paramount importance. The
't~Hooft model for two-dimensional QCD is a source of such hints. 

Thus let the meson $B$ be the pion. We substitute the explicit form of the pionic 
vertex
(\ref{pivert1}) into expression (\ref{ABC}) and after simple algebraic exercises
arrive at a striking conclusion that \cite{decay}
\be
M(A\to\pi+C)\equiv 0
\label{adlerzero}
\ee
in the chiral limit. Note that this result could be anticipated in view of the
identification (\ref{pivert}) and the current conservation laws (\ref{averv}) and
(\ref{avera}). The condition (\ref{adlerzero}) is nothing but the two-dimensional
analogue of the celebrated Adler selfconsistency condition for amplitudes 
with soft pions involved \cite{Adler}. It is not surprise, that it holds true for any value of the
pion momentum as, in view of the dimensionlessness of the pion decay constant 
$f_{\pi}$, there is no soft scale in the model which could play the role of the
edge for this condition. 

Let us conclude with a couple of comments concerning the condition (\ref{adlerzero}).
First of all, the pion turns out sterile, at least in the subleading order in $N_C$.
On the other hand, applying the above qualitative analysis concerning the dimension
of $f_{\pi}$, one can extend the condition (\ref{adlerzero}) to the case of any hadronic
process with pions involved, so that the pion is completely decoupled from the spectrum in
the chiral limit in all orders in $N_C$.

The formal reason for the condition (\ref{adlerzero}) is a totally destructive
interference between the $\vph_+$ and the $\vph_-$ components of the pionic wave function.
They were found to be of the same order of magnitude and, hence, all six terms of
the amplitude (\ref{ampl}) are equally important in establishing the condition 
(\ref{adlerzero}). The latter observation automatically invalidates any attempts to
describe the pion in the framework of a constituent quark model when the $\vph_-$
component is completely lost. It seems quite unprobable that the above
drawback of the
quark models in QCD$_2$, as well as in QCD$_4$, could be cured by simple prescriptions,
like multiplying the naive 2-term amplitude by extra \lq\lq magic"
factors \cite{Swanson}
or whatever.

\section{Conclusions and prospectives}

Phenomenological successes of quark models do tell us
that the constituent quarks are the correct degrees of freedom in the
nonperturbative domain. In these models 
the lowest $^1S_0$ $q \bar q$ state exists on the same
footing as other mesons. As it was already mentioned, there are no direct
indications that the confinement and the chiral symmetry breaking are interrelated
in case of QCD$_4$. Nevertheless, if such a scenario does not take
place, then one easily runs into troubles with constituent models:
two pions exist, one is a quark bound state, and the other one is
responsible for the chiral symmetry breaking. Of course, a roundabout way is
to disregard quark models completely. However, it is more economical to
organize the confinement and the chiral symmetry breaking due to one and 
the same mechanism.

Developing such an approach for QCD$_4$ is not a straightforward task. 
Models \cite{Orsay,port} incorporate the main ingredients,
the gap equation similar to (\ref{gap}) and the Bethe-Salpeter equation
similar to (\ref{BG}). The existence of the chirally-noninvariant solution of
the gap equation implies the existence of the Goldstone boson as the lowest
$q \bar q$ state. The $\varphi_+$ and $\varphi_-$ components of its wave
function are equal to each other in the mesonic rest frame, the axial-vector
current is conserved in the chiral limit, and all the relations of the current
algebra are satisfied. In the meantime, the role of the $\varphi_-$ component of
the wave functions for all other mesons is less drastic, and parity
degeneracy is restored for higher quarkonia \cite{Orsay}. 

Unfortunately, all these nice features
persist at the price of confining interaction employed, chosen as the time
component of a vector force. First, such a model is not covariant, that
prevents from proceeding further along the lines described above. In
particular, one cannot make Lorentz boosts within this model and
cannot describe the strong decays. Another drawback is even more important --- 
namely,
the interaction is not compatible with the area law for the isolated Wilson loop. 
This point holds true
not only for the model \cite{Orsay} where the interaction potential is the
three-dimensional oscillator one. It is also so for rather sophisticated models 
with linear confinement too (see, {\it e.g.}, \cite{Swanson2}). The reason is that 
the area law
yields linear confining potential only for heavy constituents. Besides that, the last,
but not the least, objection is the lack of the gauge invariance in such kind
of models.

An approach suggested in \cite{Sim2} is rather promising in all these
respects. The confining interaction employed there is given by a set of
gluonic field-strength correlators, which produce the area law. 
These correlators are Lorentz 
covariant by construction, and gauge invariance is preserved too. The
latter is ensured by using the generalized Fock-Schwinger gauge (Balitsky
gauge), which leads to the gluonic correlators explicitly dependent on the
reference contour and, as a consequence, explicitly translationally
noninvariant. A simple two-dimensional example of such a correlator 
is given by the expression (\ref{K}). The problem is very technically
involved due to this fact, but at the same time it is very physically transparent. 
Indeed, the interaction of such a kind describes the string which is 
developed between the constituents (for the recent progress in this
direction see \cite{SimTjon}). On the other hand, the same correlators are 
responsible for the chiral condensate formation \cite{Sim2}. We expect
that the quark model followed from such a formalism could be able to
describe $q \bar q$ bound states including the pion. 

In conclusion, we state once more that a reasonable model in four
dimensions is welcome    
in order to find the pionic solution playing the twofold role: being a
true $q \bar q$ state it is also the Goldstone boson. In our opinion the
two-dimensional 't Hooft model gives a brilliant example of how this could
happen in Nature.
\bigskip

The authors would like to thank Yu.A.Simonov for encouraging them to write this review and
for stimulating discussions, P.Bicudo and J.E.Ribeiro for the help in studying the
vacuum properties of the theory and many useful and enlightening discussions, as well as
P.Maris for reading the manuscript and critical remarks. They are also grateful to
the staff of the Centro de F\'\i sica das Interac\c c\~oes Fundamentais (CFIF-IST)
for cordial hospitality during their stay in Lisbon.

Financial support of RFFI grants 00-02-17836 and 00-15-96786,
INTAS-RFFI grant IR-97-232 and INTAS CALL 2000, project \# 110
is gratefully acknowledged. One of the authors (A.N.) is also supported by RFFI grant 
01-02-06273.
\setcounter{equation}{0}
\section{Appendix A}
\renewcommand{\theequation}{A.\arabic{equation}}

In this appendix we collect some formulae useful for the matrix approach.

The dressed quark Green's function (\ref{S}) can be rewritten in a more
convenient form if the projectors (\ref{projectors}) are introduced,
\be
S(p_{\mu})=\frac{\Lambda_+(p)\gamma_0}{p_0-E(p)+i\varepsilon}+
\frac{\Lambda_-(p)\gamma_0}{p_0+E(p)-i\varepsilon}.
\label{SS}
\ee

As mentioned before, one should be careful with the sign of the
dispersive law $E(p)$ and keep the product $\varepsilon E(p)$, when combining
the two fractions in (\ref{SS}) together \cite{Bars&Green}. The propagator
(\ref{SS}) contains all radiative corrections and should be assigned to the internal quark lines 
(fat lines in diagrams).

We find it useful to split the matrix wave function (\ref{phibig}) 
into two pieces by means of the projectors (\ref{projectors}),\footnote{In this
appendix we suppress the index numerating the mesonic states, keeping it only when
necessary.}
\be
\Phi_{\eta}(p,P)=\Lambda_{\eta}(p)\Phi(p,P)\Lambda_{\bar\eta}(P-p),\quad
\Phi(p,P)=\Phi_{\eta}(p,P)+\Phi_{\bar\eta}(p,P),
\ee
where either $\eta=+,\;\bar\eta=-$, or, {\it vice versa}, $\eta=-,\;\bar\eta=+$. 
Then the completeness condition for $\Phi$'s reads
\be
\sum_nSp\left[\Phi_+^{n+}(p,P)\Phi_+^{n}(q,P)-\Phi_-^{n+}(p,P)\Phi_-^{n}(q,P)\right]=2\pi\delta(p-q),
\ee
which can be easily derived, using the corresponding properties of the functions
$\vph_{\pm}$ and the relation
$$
Sp\left[\Phi_{\eta}^{n+}(p,P)\Phi_{\eta'}^{m}(q,P)\right]=\vph_{\eta}^n(p,P)
\vph_{\eta'}^m(q,P)\left[\delta_{\eta\eta'}\cos\frac{\theta(p)-\theta(q)}{2}
\cos\frac{\theta(P-p)-\theta(P-q)}{2}\right.
$$
\be
-\left.\delta_{\eta\bar{\eta}'}
\sin\frac{\theta(p)-\theta(q)}{2}\sin\frac{\theta(P-p)-\theta(P-q)}{2}\right].
\ee

We also give two formulae useful for the derivation of the quark-quark scattering
amplitude,
\be
\Phi^{-n}_{\eta}(p,P)=\eta_n\Phi_{\bar\eta}^{n+}(P-p,P),
\ee
\be
Sp\left[\Phi^{n+}(P-p,P)\Phi_{\eta}^m(P-q,P)\right]=\eta_n\eta_m
Sp\left[\Phi^{n+}(p,P)\Phi_{\eta}^m(q,P)\right],
\ee
where $\eta$ and $\bar\eta$ are defined above and $\eta_n$ is connected to the spatial 
parity of the state (see equation (\ref{prop2})).

The next two formulae come out directly from the bound-state equation (\ref{BG}),
\be
\Lambda_{\eta}(p)\gamma_0\Gamma(p,P)\Lambda_{\bar\eta}(p-P)\gamma_0=
\frac{1}{\gamma}\left[E(p)+E(P-p)-\eta P_0\right]\Lambda_{\eta}(p)\Phi(p,P)
\Lambda_{\bar\eta}(p-P)
\label{rel1}
\ee
$$
=\frac{1}{\gamma}\left[E(p)+E(P-p)-\eta P_0\right]\Phi_{\eta}(p,P),
$$
\be
\Lambda_{\eta}(p-P)\gamma_0\bar{\Gamma}(p,P)\Lambda_{\bar\eta}(p)\gamma_0=
\frac{1}{\gamma}\left[E(p)+E(P-p)+\eta P_0\right]\Lambda_{\eta}(p-P)\gamma_0
\Phi^+(p,P)\Lambda_{\bar\eta}(p)\gamma_0
\label{rel2}
\ee
$$
=\frac{1}{\gamma}\left[E(p)+E(P-p)+\eta P_0\right]\gamma_0
\Phi^+_{\bar\eta}(p,P)\gamma_0,
$$
whereas with the help of relations (\ref{prop1}) and (\ref{prop2}) one finds
\be
\Gamma_n(P-p,P)=\eta_n\gamma_0\bar{\Gamma}_{-n}(p,P)\gamma_0.
\ee

\end{document}